\documentstyle[12pt,amssymb]{article}
\newcommand{\msun}{M_\odot}

\newcommand{\btau}{\mbox{\boldmath$\tau$}}
\newcommand{\bpi}{\mbox{\boldmath$\pi$}}
\newcommand{\brho}{\mbox{\boldmath$\rho$}}
\newcommand{\bF}{\mbox{\boldmath$F$}}

\title{Neutron star properties with relativistic equations of state
}
\author{H. Huber, F. Weber%
, ~M. K. Weigel, and Ch. Schaab\\
Sektion Physik, Universit\"at M\"unchen\\
Am Colombwall 1, D-85748 Garching, Germany
}
\begin{document}

\baselineskip15pt
\maketitle
\begin{abstract}
We study the properties of neutron stars adopting relativistic
equations of state of neutron star matter, calculated in the framework
of the relativistic Brueckner--Hartree--Fock approximation for
electrically charge neutral neutron star matter in
beta--equilibrium. For higher densities more baryons (hyperons etc.)
are included by means of the relativistic Hartree-- or Hartree--Fock
approximation. The special features of the different approximations
and compositions are discussed in detail. Besides standard neutron
star properties special emphasis is put on the limiting periods of
neutron stars, for which the Kepler criterion and
gravitation--reaction instabilities are considered. Furthermore the
cooling behaviour of neutron stars is investigated, too. For
comparison we give also the outcome for some nonrelativistic equations
of state.
\end{abstract}

\vskip1cm
PACs numbers: 97.60\,Jd, 04.40\,Dg, 21.65+f, 97.10.Kc

\newpage
\section{Introduction}

A necessary ingredient for solving the structure equations for (rotating)
neutron stars (NS) is the equation of state (EOS) \cite{1}.
For NSs the EOS has to cover a wide range of densities ranging from
super--nuclear densities (up to 5 to 10 times normal nuclear matter density)
in the star's core down to the density of iron at the star's surface. 
At present, neither heavy--ion reactions nor NS data are capable to
determine the EOS accurately, and the behaviour of high--density
matter is still an open and one of the most challenging problems in
modern physics, containing many ingredients from different branches of
physics. The theoretical determination of the EOS over such an
enormous range has therefore to rely mainly on theoretical arguments
and extrapolations for which no direct experimental confirmation
exists. The best one can do in such a situation is a step-by-step
improvement of the available models for the EOS. Since the central
density of a NS is so extreme, both the Fermi momenta and the
effective nucleon mass are of the order of 500~MeV, one should prefer
a relativistic description
\cite{2}-\cite{9}. 

Neutron star matter differs from the high density systems produced in
heavy ion collisions by two essential features: a) Matter in high
energy collisions is still governed by the charge symmetric nuclear
force while neutron star matter (NSM) is bound by gravity. Since the
repulsive Coulomb force is much stronger than the gravitational
attraction, NSM is much more asymmetric than ``standard'' matter. b)
The second essential difference is caused by the weak interaction time
scale of $\sim 10^{-10}$s, which is small in comparison with the
lifetime of the star, but large in comparison with the characteristic
time scale of heavy ion reactions. For that reasons ``normal'' matter
is subject to the constraints of isospin symmetry and strangeness
conservation, but NSM has to obey the constraints of charge neutrality
and generalized beta--equilibrium with no strangeness conservation. It
is obvious from these considerations that NSM is an even more
theoretical object than nuclear ``normal'' matter, with a rather
complex structure \cite{2}-\cite{9}.

Due to these features one can adopt the following structure of a
neutron star: The two outer crusts of the star have crystalline
structures, for which rather reliable EOSs exist in the literature. In
the uniform density region ($\rho \gtrsim \times
10^{14}~\mbox{g/cm}^3$) of neutron star matter, one has to deal with a
system of interacting baryons (i.e., $p,n, \Lambda, \Sigma,$ possible
$\Delta'$s, etc.) and/or quarks ($u,d,s$ flavours, uniform or not
uniform) that are in generalized $\beta$--equilibrium with leptons
($e^-,
\mu^-$) \cite{2}-\cite{11}. Furthermore one may encounter meson condensates
\cite{2,11,12}.

Despite the differences of hot symmetric non--strange matter produced
in high energy collisions and cold asymmetric charge neutral and
strangeness containing matter of NSs one should combine both systems
in a common theory. This is possible to a certain degree in modern
field theoretical relativistic approaches, and NSs are unique systems
in the sense, that they offer a test bench for the EOS of exotic NSM,
which can not be mimicked in terrestrial laboratories.

The first attempt, based on the Fermi gas model, to incorporate the role of the
hyperons in NSs is due to Ambartsumyan and Saakyan \cite{13}, which was later
improved in the nonrelativistic scheme by several authors \cite{14}. The
first systematic investigation in the relativistic framework was performed by
Glendenning, who used the relativistic mean--field approximation with
inclusion of hyperons and $\Delta$'s. \cite{8}. This model was later extended
by using the relativistic Hartree--Fock--approximation (RHF)
\cite{7,9,10} and by improving the Lagrangian in the mean--field
approximation (or relativistic Hartree--approximation; RH)
\cite{6,11,15,16}.  One of the most exhaustive investigations in the
later framework was performed by Schaffner and Mishustin, who used
modern phenomenological interactions in the nucleon and hyperon
sector, respectively. In this last model kaon condensation turned out
to be unlikely \cite{11}.

From a microscopic standpoint such treatments are not
satisfactory, since the interaction is adjusted in a phenomenological manner
to properties of finite nuclei and the parameters of the Bethe--Weizs\"acker mass
formula. Other drawbacks are, for instance, the vanishing of the
$\pi$--meson contributions in the mean--field approximation, and large
$\sigma$--meson self--interactions are needed to reduce the
incompressibility $K$. Some of these deficiencies do not occur in the
framework of the relativistic Hartree--Fock approximation, which also
resembles, in both its mathematical structure and its Lagrangian density
rather closely to the microscopic relativistic Brueckner--Hartree--Fock
theory, for which one--boson--exchange potentials (OBEP) are adjusted to
the two--nucleon data. In order to incorporate a more microscopic
description of neutron star matter, we have recently developed a model in
which this  matter is described for moderate densities by self--consistent
RBHF--calculations for $\beta$--stable matter ($n,p,e^-$ and $\mu^-$ only)
\cite{7,17}. Due to technical difficulties, RBHF--calculations are presently
restricted to densities up to 2--3 times nuclear matter density. Since
at higher densities other baryon states become populated and
RBHF--calculations with inclusion of such states are presently not
feasible, we extrapolated the EOS at higher densities within the RH
and/or the RHF approximation. The essential new feature of this scheme
in comparison with older treatments was the different adjustment of
the parametrization of the RH and RHF Lagrangians. Normally one
restricts oneself to the reproduction of the properties of symmetric
nuclear matter, but in the new parametrization we used the outcome of
RBHF calculations of asymmetric and NS--matter as well.  It turned out
that the RH approximation is not so flexible in reproducing the
properties of NSM (asymmetry, composition etc.).  Since the Hartree
approximation has been used in the vast majority of the earlier
investigations, we will include this approximation in our
considerations, too, which has also the advantage of greater
transparency.  Special attention will be paid to the
RHF--approximation, where new features enter into the properties of
NSM. In both approximations it is possible to incorporate more baryon
states (for more details, see Refs.\,\cite{4,7,9} and following
sections).

In order to achieve a certain degree of selfcontainment we
recapitulate also the basic theory for the EOSs, and for the
properties of neutron stars, i.e.  structure of rotating stars in
general relativity, stability of rotations (Kepler criterion,
gravitation--radiation instabilities), cooling of neutron stars
etc. The contribution is organized as follows. Section\,II is devoted
to the EOS of NSM, where we discuss, after a brief theoretical review
the different models of the EOSs. The next section deals with the
properties of NSs and the summary is given in Section\,IV.

\section{Equation of state}
\setcounter{equation}{0}
\noindent
{\bf a) General theory:}

\vskip1ex
The EOSs of neutron star matter were determined as described in
Refs.\,\cite{17,18}. For the two outer crusts we used EOSs taken from the
literature \cite{19,20}. The general dynamics of the hadron/lepton system in
the uniform
region of neutron star matter is governed by an OBE Lagrangian of the
following form:
\begin{eqnarray}
\label{1}
{\cal L}(x)&  = &  \sum_{B = p,n,\Sigma^{\pm
0},\Lambda,\Xi^{0,-},\Delta^{-,0,+,++}} {\cal L}^0_B (x) \\ \nonumber
& & +~\sum_{M=\sigma,\omega,\pi,\rho,\delta} \left\{
   {\cal L}^0_M (x) + \sum_{B=p,n,\ldots\Delta} {\cal L}^{int}_{B,M} (x)
   \right\} + \sum_{L=e^-,\mu^-} {\cal L}_L (x) ~.
\end{eqnarray}
The forces between the different baryons are mediated by the exchange
of different mesons $M=\sigma,\omega,\ldots$\. The leptons $e^-$ and
$\mu^-$ are treated as free particles. Furthermore one has to impose
the constraints of beta--equilibrium ($q_B$ and $\mu_B$ denote the
electric charge and chemical potential of the baryon $B$,
respectively):
\begin{equation}
\label{2}
\mu_B = \mu_n - q_B \mu_e  \qquad , \qquad \mu_\mu = \mu_e ~.
\end{equation}
and charge neutrality.  For the many--body treatment we employed the
relativistic many--body Green's function scheme. Here one has to solve
-- for Brueckner-type approximations -- a coupled system of equations
which consists of the Dyson equation for the Green's function $G$, the
effective scattering matrix $T$ in matter, and the equation for the
self--energy $\Sigma$ of a baryon in matter \cite{21}:
\begin{eqnarray}
\label{3}
{(G^0)^{-1}(1,2) \Sigma(1,2)} G(2,1') & = & \delta(1,1') \\
\label{4}
<12|T|1'2'> & = & <12|V|1'2' - 2'1'> + \\ \nonumber
            &   & i<12|V|34> \Lambda(34,56) <56|T|1'2'>
~, \nonumber 
\end{eqnarray}
with
\begin{equation}
\label{5}
\Sigma(1,2)  =  - <14|T|52> G(5,4) ~.
\end{equation}

For the intermediate baryon--baryon propagator $\Lambda$, we have chosen in
the RBHF approximation the Brueckner propagator. $V$ denotes the OBE
potential, which has the following structure
\begin{equation}
\label{6}
<12|V|1'2'>~ = \sum_{M=\sigma,\omega,\rho\ldots} <12|V_M|1'2'> ~.
\end{equation}
The RH and RHF approximations are obtained for $T = V$ and $T = V -
V^{ex}$\,, respectively. For the OBE potentials we selected the
potentials A and B constructed by Brockmann and Machleidt
\cite{22}. In the RBHF--treatment, performed in the full Dirac space,
of symmetric and asymmetric nuclear matter both potentials (they
differ in the strength of the tensor force, which increases from A to B)
give good agreement with the nuclear matter parameters ($E/A,
\rho_{00}, K_v, J$), and also the volume parameters $L$ and $K_{sym}$ of
asymmetric matter are in accordance with the data. An illustrative comparison
with other treatments is given in Table\,I \cite{21}.

Up to 2--3 times equilibrium nuclear matter density we used for the
EOSs the outcome of an RBHF--calculation for neutron star matter by
restriction to $p,n,e^-,$ and $\mu^-$, only.  For higher densities
more baryon states $B$ become populated in $\beta$--stable neutron
star matter. For high density systems with such a complex composition
the RBHF scheme is not feasible. In order to find an extension of the
microscopic RBHF--EOS to higher densities we selected both the
relativistic Hartree and the relativistic Hartree--Fock approximation,
where the parameters in the nucleonic sector (coupling constants etc.)
are adjusted to the results of RBHF--calculations for asymmetric and
NS matter.

Explicitly, the Lagrangian density for the RHF--approximation, where the
forces are mediated by the exchange of $\sigma$--, $\omega$--, and
$\rho$--mesons, and $\sigma$ self--interactions are included, is given by
\cite{17}:
\begin{eqnarray}
\label{II.6n}
\lefteqn{
        {\cal L}(x) =  \sum_B \bar\psi_B(x) \left[
        i\gamma^\mu \partial_\mu - m_B + g^B_\sigma \sigma(x) - g^B_\omega
        \gamma^\mu \omega_\mu(x) - f^B_\omega \frac{\sigma^{\mu\nu}}{4m_B} 
        F^\omega_{\mu\nu} \right.
} \nonumber \\
&&\left. -\,\frac{f\pi}{m\pi}\,\gamma^5 \gamma^\mu \btau_B \cdot
\partial_\mu\bpi 
   - g_\rho^B \gamma^\mu \btau \cdot \brho_\mu(x) -
   f^B_\rho \frac{\sigma^{\mu\nu}}{4m_B} \btau \cdot \bF^\rho_{\mu\nu}\right]
  \psi_B(x) \nonumber \\
&& +\,\frac{1}{2} \left[\partial_\mu \sigma(x) \partial^\mu \sigma(x) -
  m^2_\sigma \sigma^2(x)\right] +
  \frac{1}{2} \left[ \partial_\mu \bpi(x) \cdot \partial^\mu \bpi(x) -
  m^2_\pi \bpi^2(x)\right] \nonumber \\
&& -\, \frac{1}{4}\bF^\rho_{\mu\nu}(x) \cdot \bF^{\mu\nu,\rho}(x) +
  \frac{1}{2} m^2_\rho \brho^{\mu}(x) \cdot \brho_\mu(x) -
  \frac{1}{4} F^\omega_{\mu\nu}(x) F^{\mu\nu,\omega}(x) +
  \frac{1}{2} m^2_\omega \omega^\mu(x) \omega_\mu(x) \nonumber \\
&& -\,\frac{1}{3} m_N b_N [g_\sigma \sigma(x)]^3 - \frac{1}{4} c_N [g_\sigma
\sigma(x)]^4 ~,
\end{eqnarray}
with
\begin{equation}
\label{II.7n}
 F^\omega_{\mu\nu}(x)
   \equiv \partial_\mu \omega_\nu(x) - \partial_\nu \omega_\mu(x) ~, \quad
\bF^\rho_{\mu\nu}(x) \equiv \partial_\mu \brho_\nu(x) - \partial_\nu
\brho_\nu(x)~. 
\end{equation}
With respect to the saturation properties of infinite nuclear matter
(INM) (see Table\,I) the results are identical to the RBHF--outcome.  It
seems that, at present, such a procedure is the only possibility to
incorporate more baryons and to establish a connection to microscopic
RBHF--calculations with realistic OBE--potentials simultaneously. In
this context one has to remark that the standard relativistic approach
in almost all cases is based on a pure phenomenological RH--treatment,
where the parameters in the nucleonic sector are adjusted to the
properties of symmetric nuclear matter, so that our approach has the
advantage to contain more microscopic elements (for more details, see
Refs.\,\cite{17,18,21}).  In the calculations we used an improved
parametrization \cite{18}, and the parameter sets are given in
Table\,II.

\vskip2ex
\noindent
{\bf b) Comparison of the different approximations}

\vskip1ex
First one has to test whether the described approximations can
reproduce the properties of asymmetric and NS matter in the density
range, where the RBHF--treatment is applicable (nucleons and leptons
only). In Fig.\,1 we compare the EOSs for different asymmetries in the
different approximations.  The agreement of the RBHF--EOSs with the
RHF--EOSs is rather satisfactory for the whole asymmetry
range. However, for the RH--EOSs the agreement is worse for larger
asymmetries, furthermore the RH--EOS becomes stiffer for higher
densities.  This behaviour of the RH--approximation can better be
inferred from Fig.\,2, where the comparison for the pressures for NSM
is shown. In Fig.\,3 we exhibit, as a further example, the comparison
with respect to the baryon/lepton composition. As discussed in
Refs.\,\cite{17,18} the RHF--approximation with no $\rho$--tensor
coupling gives the best agreement with the RBHF--EOS. The
RHF--approximation has the additional advantage that the Lagrangian
density and the mathematical structure resembles more the RBHF--scheme
than the RH--approximation (no $\pi$--meson, exchange contributions
etc.). For the presentation we selected the Brockmann--Machleidt
potential $B$\,\cite{22}; for the potential $A$ the situation is
completely analogous \cite{17,18}.

\vskip2ex
\noindent
{\bf c) Neutron star matter}

\vskip2ex
\noindent
$\alpha$) General considerations 

\medskip
For the calculation of NS--properties the EOS is needed for a wide range of
densities, stretching to several times of nuclear matter saturation density
\cite{2}-\cite{7}. If one extrapolates the described scheme to the density
domain of NSM, one faces the following dilemma: It is well known that
NS--properties depend strongly on the properties of the EOS near
saturation, which is obvious for lighter NSs but also true for heavier
stars \cite{26}.  Therefore one should use in this region an EOS,
which is either based on microscopic RBHF--calculations
\cite{21,22,27,28} or phenomenological parametrizations of the
RH/RHF--approximation, adjusted to nuclear data \cite{23}. In both
cases one obtains inevitably a rather stiff EOS caused by the low
value of the Dirac mass of about 0.6~$m_N$ (correct spin--orbit
splitting \cite{29}; the RBHF--scheme gives also reasonable results
for finite nuclei \cite{30}) at saturation \cite{23,31,32}. The
resulting meson fields are then rather large and consequently one
obtains a sharp drop of the Dirac masses with increasing density. This
feature is even amplified by the unavoidable occurence of negative
values for $c_N (b_N/c_N \sim -1)$ \cite{23,31,32}, which causes a
nonmonotonic behaviour of the effective $\sigma$--mass
\begin{equation}
\label{II.9n}
m^{*2}_\sigma = m^2_\sigma + g^2_\sigma \left[ b_N m_N <g_\sigma \sigma> +
   c_N <g_\sigma\sigma>^2 \right]~,
\end{equation}
which increases the attraction beyond $<g_\sigma\sigma>^0 \equiv -
b_Nm_N/2c_N$. As long as the composition of NSM is restricted to
$n,p,e^-$, and $\mu^-$ only, the EOS is sufficiently stiff to reach
the necessary central pressure of the star at moderate densities (see
Section\,III.e). However if one includes more baryons in the
NSM--composition the EOS becomes considerably softer and higher
densities are needed to obtain sufficient central pressure. The scalar
fields are therefore in this case rather large and hence negative
Dirac masses for the nucleons occur in the calculation
\cite{11,18}. One has tried to overcome this problem in a
phenomenological manner using so--called stabilized
$\sigma$--functional forms \cite{23,33}, for which the dangerous
negative curvature of $m^{*2}_\sigma$ is switched off. However for the
familiar parametrizations PL--2 and PL--40 negative Dirac masses also
still occur in NSM.  Schaffner and Mishustin have circumvented this
problem by the rather questionable ad hoc assumption of using always
absolute values for the Dirac masses, so implicitly changing the
stiffness of the EOS \cite{11}.  Another approach by which one gets
now along with a reduced attraction, i.e. $b_N, c_N > 0(c_N \gg b_N)$
reduces the repulsion by an additional $\omega$--self--interaction
\cite{34,35}. In this manner one obtains softer EOS with smaller
$\sigma$--fields, where negative Dirac masses do not occur. However
the resulting EOS may be too soft and an additional repulsion among
the hyperons seems necessary \cite{11}. Furthermore the asymptotic
behaviour is changed from the standard behaviour proportional to
$\rho^2_B$ to that of an ideal gas ($\propto
\rho_B^{4/3}$). Unfortunately the NSM--results for this case were not
applied to NS--properties in Ref.\,\cite{11}. For the sake of
comparison we performed therefore some NS--calculations and obtained
-- as expected -- maximal masses around 1.5~M$_\odot$. (For instance,
for the parameter set TM\,1 we obtained 1.561\,M$_\odot$ with a
central energy density $\epsilon_c = 740$~MeV/fm$^3$ \cite{18}.) One
might wonder why this problem was not discussed in the other
investigations of NSM. In the vast majority one prefers an EOS which
is based on the standard nuclear matter parameters but with a Dirac
mass of approximately 0.78~$m_N$
\cite{6,8,15,16,26}. Only in this window \cite{31,32} one obtains positive
values for $c_N$ (or $b_N \gg|c_N|)$ and more moderate meson fields,
so that the problem of negative nucleon Dirac masses is avoided. The
reasons given for this choice rest on a reproduction of the effective
mass ($\sim 0.83~m_N$) \cite{8,36}, however a closer inspection
according to a more elaborate investigation by Celenza and Shakin
shows that values of again 0.6$~m_N$ are more appropriate for the
Dirac mass \cite{37} (see also Ref.\,\cite{38}). Furthermore one
favours rather low values for the saturation density and higher values
for the incompressibility in order to stiffen the EOS (see, for
instance, Ref.\,\cite{8}). For instance, according to the
Hugenholtz--vanHove theorem, $g_{\omega N}$ varies by approximately
40\% in the range of $1.3~\rm{fm}^{-1}\leq p^0_{F}\leq 1.42~\rm{fm}^{-1}$
\cite{31}.

Common to all these approaches is to impose a certain behaviour of the
NSM--EOS in the NSM--domain, which is not known. Controlled is this
behaviour by an additional parameter (large Dirac mass; switching off
parameter; coupling constant of the vector self--interaction etc.
\cite{6,8,16,23,33,35}). In order to overcome this dilemma, namely to keep
the
connection to the EOS in the vicinity of nuclear saturation and the described
unpleasant features of the NSM--EOS with additional baryons one is also
forced in our
approach to implement a working extrapolation hypothesis controlled by an
additional parameter. Also RBHF--calculations in symmetric matter favour
higher Dirac masses in this density region \cite{38}. In order to keep the
scheme as simple as possible we extrapolate as follows: We maintain the
dynamics till the maximum of the effective $\sigma$--mass and extrapolate
beyond this point via a Lagrangian of the same structure, but with new
self--interaction couplings and a modified $\sigma$--mass:
\begin{equation}\label{II.10n}
m^2_\sigma \to m'^2_\sigma = m_\sigma^2 + \frac{g_\sigma^2}{4c_N} m^2_N b^2_N 
     \left( \frac{\alpha}{c_N} - 1 \right) ,
\end{equation}
\begin{equation}\label{II.11n}
b_N \to b'_N = \alpha \frac{b_N}{c_N} \qquad ; \qquad c_N \to c'_N = \alpha
~.
\end{equation}
For this model the effective $\sigma$--mass and its derivative agree
with the microscopic model at the transition point (see
Eq.\,\ref{II.9n}). $\alpha = c_N$ gives the original dynamics;
$\alpha= 0$ results in a linear Walecka model, which turns out to be
insufficient \cite{18} (negative Dirac masses). With $\alpha > 0$ one
can now control the stiffness.  One should remark in this context that
the main effect of the extrapolated dynamics is a more moderate drop
of the Dirac masses beyond four times nuclear matter saturation
density in accordance with Ref.\,\cite{38} (see Fig.\,4). With respect
to the pressure, the differences against the original dynamics are
rather small (10\%) in the lower part of the NSM--density domain and
become very small in the high--density NSM--region, where the
unchanged $\omega$--meson repulsion dominates.  We have also tested
the pure linear or quadratic extrapolations, i.e.  $b'_N = b_N/2, c'_N
= 0$ \,or\, $b'_N = 0, c'_N = -c_N;~ m'_\sigma = m_\sigma,$ but their
results can be incorporated closely in the scheme selecting special
$\alpha$--values \cite{18}.
\vskip2ex
\noindent
$\beta$) Results and discussion

\medskip
Essentially three main features characterize the properties of NSM: a) the
 stiffness of the EOS  b) the relative coupling constants of hyperons,
c) the chosen many--body approximation. All three points are correlated, but
nevertheless we will try to separate them to a certain degree in order to
extract some insight into the structure of the problem.

The influence of the first point is obvious, since the necessary central
pressure for stable stars can be reached earlier for larger stiffnesses.
Important is the second issue. The relative hyperon couplings are not well
known, since the information from hypernuclei data (for more details, see,
Refs.\,\cite{39,40}) permits a wide bandwidth (for instance, $0.4\leq
\mbox{x}_{H\sigma} \equiv g_{H\sigma}/g_{N\sigma}\leq 0.8$ in the relativistic
mean field approximation). Therefore a number of choices were made in
the literature, reaching from universal coupling, which gives a first
insight, to ratios motivated by the quark model
\cite{4,6,7,8,9,11,15,16}. A suitable first choice in the latter case
is to use the SU(6) symmetry for the vector couplings i.e.,
\begin{equation}
\frac{1}{3}g_{\omega N} = \frac{1}{2}g_{\omega \Lambda} =
\frac{1}{2}g_{\omega
\Sigma} = g_{\omega\Xi}~;~~g_{\rho N} = g_{\rho\Sigma} =
g_{\rho\Xi}~,~~g_{\rho\Lambda} = 0,
\end{equation}
and to fix the $\sigma$--coupling according to the potential depth of the
hyperon--particle in nuclear matter ($\sim - 30~$MeV). According to these
uncertainties we will investigate the problem in a more systematic manner by
making several choices for the $\sigma$--hyperon coupling and fix the
$\omega$--coupling by means of the hyperon potential depth
\cite{15,40}. The selected choices  for the EOSs are described in
Table\,III. The hyperon couplings are given in Table\,IV.

The general trend is that the EOSs become softer with decreasing
couplings, because then the conversion of nucleons into hyperons is
energetically favourable, due to the smaller repulsive force
dominating in the high--density domain of NSM. A special feature of
the Hartee approximation are large $\rho$--couplings, which are
necessary for the adjustment of the symmetry energy \cite{17}. For
that reason the charge--favoured (isospin unfavoured) $\Delta^-$ does
not occur (or play a minor role; see discussion of the
RHF--approximation) and therefore is usually neglected a priori in
this approximation \cite{6,11,15,16}. For the $\Sigma^-$ the charge
compensation still dominates and so the $\Sigma^-$ occurs rather
early. Increase of the $\rho$--coupling, for instance like
$g_{\rho\Sigma}= 2g_{\rho N}$, would reverse the onset of $\Sigma^-$
and $\Sigma^+$ \cite{18}. A further and more severe point is the
selection of the many--body approximation. For the Hartree
approximation we obtain compositions and stellar properties which are
familiar to investigations performed earlier in this framework. This
approximation is characterized as a high density approximation with a
relatively stiff EOS containing no exchange contributions. For the
latter reason the coupling constants are larger than in the
RHF--scheme.  Two examples for the composition (universal coupling and
SU(6) scheme) are given in Figs.\,5 and 6, from which one can infer
the dependence on the hyperon couplings. For the universal coupling
the attraction is more favoured, which reduces the Dirac
masses. Therefore the onset of the hyperons occurs earlier and also
$\Delta$'s are possible. This causes a lower pressure at lower
densities, however for high densities the stronger repulsion causes
then a stiffer EOS (see Fig.\,7). For these reasons one should obtain
smaller NS--masses for the universal coupling for smaller central
densities than for the SU(6)--coupling. For higher densities the
situation is reversed (see Sec.\,III.e).

Due to the additional degrees of freedom the RHF--EOS is softer than
the corresponding RH--EOS in a large density domain.  A special new
feature of the RHF--approximation is that the onset of a particular
baryon species in NSM is, as in the RH--approximation, solely
controlled by the Hartree term. Due to the lower couplings the
participation of the other baryons is more enhanced and in general the
onset of the hyperons occurs earlier than in the RH--scheme, resulting
in a quite different baryon/lepton composition and a softer EOS.  A
further characteristic difference is caused by the smaller
$\rho$--coupling constants in the RHF--theory (see
Table\,II). Responsible for this decrease of the coupling constants
$g_\rho$ are -- as in the case of realistic OBE--potentials in the
RBHF--theory -- the exchange contributions, which contribute to the
symmetry energy. For that reason the charge--favoured but isospin
unfavoured $\Delta^-$ plays now -- as in former nonrelativistic
many--body approximations with correlations \cite{14} -- an important
role in the composition.  One might further wonder that one has not
included the $\omega$--tensor coupling for the hyperons, since, for
instance, according to the quark model
$f^\Lambda_{\omega}/g^\Lambda_{\omega}$ becomes $-1$ for the $\Lambda$
hyperon. For that reason we included this term in a test calculation
\cite{18} but the impact on the EOS was rather small and can therefore
be neglected in comparison with the other uncertainties.

If one would use naively the same hyperon couplings as in the RH--approach
one obtains compositions -- shown for the SU(6) in Fig.\,8 -- where the
$\Delta$'s occur rather early (for universal coupling the baryons occur in
the order $\Sigma^-, \Lambda, \Delta^-$ with a high contribution of $\Delta$'s
at higher density \cite{18}). For that reason the EOSs are rather soft at
moderate densities and should cause a weak increase of the NS--mass in the
lower part of the central energy--density region resembling in this part the
behaviour of NS--masses calculated with EOSs with rather low incompressibilities \cite{41,42} (see
Fig.\,17). The strong abundance of the $\Delta$'s in such EOSs is caused, as discussed before, by the early onset of the charge-favoured $\Delta^-$, which is not so strongly isospin hindered in the RHF-approximation. The onset of hyperons in this model may be overestimated, since it is
controlled by their Hartree contribution solely. A special feature of the
RHF--theory seems to be that the Fock contributions for the scalar part of
the self--energy approximately cancel each other but for the vector part they
amount to 50\,\% of the total value \cite{18}. For that reason the hyperon
couplings taken from the RH--approach do not reproduce the potential depths
of the hyperons in nuclear matter in the RHF--scheme (attraction too strong)
and hence favour the hyperons. If one now corrects this deficiency by
adjusting the hyperon couplings ($x_{\sigma H}$ (RHF) $< x_{\sigma H}$ (RH);
see Table\,IV), the composition becomes now dominated by the $\Delta$'s (see
Fig.\,9). The occurence of the hyperons ($\Lambda$--hyperon) is rather late
even for weak couplings (for instance, for $x_{\omega H}$ = 0.5 at $\rho
\sim 0.5$~fm$^{-3}$). The peculiar behaviour of the pressure disappears
now \cite{18} and consequently the NS--masses as function of the central
energy--density give now the standard strong increase for moderate densities
(see Fig.\,19, 20). The models treated so far involve the assumption that the
$\Delta$--coupling agrees with the nucleon coupling. But according to
investigations of ter Haar and Malfliet \cite{43}, the $\Delta$--mass does
not change very much from the vacuum case, indicating smaller
$\Delta$--couplings, which would delay the onset of the $\Delta$s. In a
recent treatment by R.~Rapp et al. \cite{44}, the choice $g_{\Delta\Delta} =
g_{\Delta N} = 0.625~g_{NN}$ was recommended, but one could use as an option
also this choice for the attraction only \cite{14}. For the first case one
gets a slight shift of the $\Delta$-threshold towards higher densities
(see Fig.\,10) and the EOS becomes much softer due to the lower repulsion.
This can be inferred from Fig.\,20, where the maximum star mass is
approximately 1.5~M$_\odot$. In the last case one deals with a strong
repulsion combined with moderate
masses for the $\Delta$'s. For that reason one expects a minor role to be played by
$\Delta$s. The calculation of the chemical potentials shows that the
attraction is just a little bit to small for the occurence of the $\Delta$'s
and one gets a pattern familiar from the RH--treatment (see Fig.\,11).
  If one reproduces the same $\Delta$--potential
depths as in the RH--scheme, the
$\Delta^-$ is still preferred at lower densities,
but hyperons play now a more significant role due to their lower repulsion
and dominate the high density region (see Fig.\,12). With respect to the
adjustment of the potential depths of the baryons, this RHF--model
corresponds exactly to the treatment within the RH--scheme.
As in the case of nucleons, the different assumptions about the behaviour of the
$\Delta$--Dirac mass in matter influence the EOS only weakly. For instance,
in the SU(6)--scheme for the hyperons the pressure increases only slightly by
going from RHF\,8 via RHF\,9 to RHF\,1\,. More severe are changes of the
repulsion of the baryons. For example, one obtains by increasing the
hyperon repulsion by going from RHF\,1 to RHF\,4 ($x_{\omega H}=0.8$)
pressure increases of $\sim$150~MeV/fm$^3$, compared with $\sim$
30~MeV/fm$^3$ in the case before with constant hyperon couplings \cite{18}.

In conclusion we have constructed and discussed EOSs of NSM, which,
in contrast to former investigations, are tight to the outcome of microscopic
RBHF--calculations. The extrapolation to higher densities, where more baryons
participate, was  performed in the RH-- and RHF--scheme. For
the density domain of NSM, where the EOS is (completely) unknown, we were
confronted as in all treatments with a complex composition with the necessity
to invoke assumptions about the behaviour, especially with respect to the
density dependence of the Dirac masses, controlled by an additional
parameter. In this context we would like to emphasize again that so far
almost all relativistic models have considered the rather
special case of a RH--EOS based on  a large Dirac mass at saturation ($\sim
.79~m_N$). With respect to the couplings of the hyperons and $\Delta$'s large
uncertainties exist. For the hyperons the relative ratios of the $\sigma$\,--
and $\omega$\,--\,couplings were fixed by the potential depths in nuclear
matter.
(An absolute fixing from hypernuclei data is not possible at present
\cite{40}.) For
the $\Delta$--coupling we invoked several assumptions, which take into
account the smaller decrease of the $\Delta$--mass in matter.

The case of the RH--treatment resembles in its basic features to
earlier investigations in this framework, which were based on more or
less phenomenological treatments. Their distinguishing features are a
high hyperon content (high strangeness), stiffer EOSs, and a strong
$\Delta$--suppression. In the RHF--scheme the $\Delta$'s play in
general due to their smaller $g_\rho$--coupling an important role, and
the hyperons are more suppressed. The detailed composition and
stiffness of the EOS depends on the Dirac mass of the $\Delta$, for
which we selected four different choices.

We are aware of the fact that the described uncertainties open the gates to a
realm of options, some of which might be rather unfamiliar in comparison with
standard treatments. Of course we could not explore and present all the
hypotheses, but we have tried to select illustrative samples, from which the
trends can be extracted (for more details, see Ref.\,\cite{18}). The
consequences for NSs will be discussed in the following sections.

\section{Neutron star properties}
\setcounter{equation}{0}

\noindent
{\bf a) General considerations}

\vskip1ex
The theoretical description of a neutron star is governed by the following
conditions, which can be roughly estimated, for instance, on the basis of the
Fermi gas or the hard core model for the EOS: i) General Relativity has to be
taken into account for the determination of the gross properties of a star
with approximately one solar mass $M_\odot$ and a radius $R$ of approximately
10\,km, since the relativistic effects (change of the metric etc.) for such
objects are of the order \cite{1}
\begin{equation}
\label{2.1}
\frac{2M}{Rc^2} \sim 0.3 ~.
\end{equation}
ii) For the EOSs one can use relativistic treatments within a
Minkowski metric since the spacing of baryons in the star ($R = r_0
A^{1/3}; r_0 \sim 0.5~\rm{fm}, A \sim 10^{57}$) is of the order of
$10^{-19}$. iii) As explained before, due to the high densities one
should favour relativistic EOSs for NSM.

Roughly one can cast the EOSs in terms of stiff or soft equations. The
``stiff'' equations give a maximum mass (Oppenheimer--Volkoff limit
M$_{OV}$) of about twice the solar mass, and limiting rotational
periods larger than 1.5~ms.  For ``soft'' EOSs the estimates are
M$_{OV}$ = 1.5~M$_{\odot}$ and P$_{min}$ = 0.6ms. Hence, the detection
of sub--milliseond or heavy pulsars could discriminate the
(theoretical) EOSs. Furthermore the millisecond pulsars, explained
according to the present understanding within the so--called recycling
model, are very fast rotating objects (at birth at least 0.5~ms)
\cite{45}. Consequently it is important to include the (fast) rotation into
the theoretical treatment of NSs.

\vskip2ex
\noindent
{\bf b) Theoretical treatment of rotating and deformed stars}

\vskip1ex
Due to these conditions one is faced with the problem to determine Einstein's
curvature tensor $G_{\mu\nu}$ for a massive star ($R_{\mu\nu}, g_{\mu\nu}$,
and $R$ denote the Ricci tensor, metric tensor, and Ricci scalar,
respectively).
\begin{equation}
\label{2.2}
G_{\mu\nu} \equiv R_{\mu\nu} - \frac{1}{2} g_{\mu\nu} R =
8\pi\,T_{\mu\nu}\left(\epsilon, P(\epsilon)\right)~.
\end{equation}
A necessary ingredient for solving (III.2) is the energy--momentum tensor
density $T_{\mu\nu}$, for which knowledge of the (relativistic) EOS, i.e.
pressure $P$ as function of the energy density $\epsilon$ is necessary.

For a spherically symmetric and static star, the metric has the famous
Schwarzschild form ($G=c=1$):
\begin{equation}
\label{2.3}
ds^2 = -\,e^{2\phi(r)} dt^2 + e^{2\Lambda(r)} dr^2 + r^2 (d \theta^2 + sin^2
\theta d\phi^2)~,
\end{equation}
where the metric functions are given by:
\begin{equation}
\label{2.4}
e^{2\Lambda(r)} = (1 - \gamma(r))^{-1} ~,
\end{equation}
\begin{equation}
\label{2.5}
e^{2\phi(r)} = e^{-2\Lambda(r)} = (1 - \gamma(r)) \quad \rm{for} \quad r >
R_{star}~,
\end{equation}
with
\begin{equation}
\label{2.6}
\gamma(r) = \left\{
\begin{array}{cc}
\frac{2M(r)}{r} & r<R_s \\[2mm]  \frac{2M_s}{r} & r>R_s
\end{array} \right.
\end{equation}
Einstein's equations for a static star reduce then to the familiar
Tolman--Oppenheimer-Volkoff equation (TOV) \cite{1,4,7}:
\begin{equation}
\label{2.7}
\frac{dP(r)}{dr} = -\,\frac{1}{r^2} \left(\epsilon(r) + P(r)\right)
      \left(M(r) + 4\pi r^3P(r)\right)  e^{-2\Lambda(r)} ~,
\end{equation}
where the gravitational mass $M(r)$ contained in a sphere with radius $r$
is determined via the energy--density $\epsilon(r)$ by:
\begin{equation}
\label{2.8}
M(r) = 4\pi \int^r_0 \epsilon(r) r^2 dr ~.
\end{equation}
The metric function $\phi(r)$ obeys the differential equation
\begin{equation}
\label{2.9}
\frac{d\phi}{dr} = -\, \frac{1}{\epsilon(r)+P(r)} \, \frac{dP}{dr} ~,
\end{equation}
with the boundary condition
\begin{equation}
\label{2.10}
\phi(r = R_s) = \frac{1}{2} \ln (1-\gamma(R_s)) ~.
\end{equation}
For a given EOS i.e. $P(\epsilon)$, one can now solve the TOV equation by
integrating them for a given central energy density $\epsilon_c$
from the star's centre to the star's radius, defined by $P(R_s) = 0$.

More complicated is the case of rotating stars, where due to the rotation
changes occur in the pressure, energy density, etc. The energy--momentum
density tensor $T_{\mu\nu}$ takes the form $(g^{\mu\nu}u_\mu u_\nu = -\,1)$
\cite{4,7,46,47}:
\begin{equation}
\label{2.11}
T_{\mu\nu} = T^0_{\mu 0} + \Delta T_{\mu\nu} ~,
\end{equation}
with
\begin{equation}
\label{2.12}
T_{\mu\nu}^0 = (\epsilon + P) u_\mu u_\nu + P g_{\mu\nu} ~,
\end{equation}
\begin{equation}
\label{2.13}
\Delta T_{\mu\nu} = (\Delta \epsilon + \Delta P) u_\mu u_\nu + \Delta P
g_{\mu\nu} ~.
\end{equation}
$P, \epsilon$, and $\rho$ are quantities in a local inertial frame comoving
with the fluid at the instant of measurement. For the rotationally deformed, axially--symmetric
configurations one assumes a multipole expansion up to second order ($P_2$
denotes the Legrendre polynomial):
\begin{eqnarray}
\label{2.14}
\Delta P & = & (\epsilon + P) (p_0 + p_2 P_2 (\cos \theta)) ~, \\
\label{2.15}
\Delta \epsilon & = & \Delta P \frac{\partial \epsilon}{\partial P} ~, \\
\label{2.16}
\Delta \rho & = & \Delta P \frac{\partial \rho}{\partial P} ~.
\end{eqnarray}
For the rotating and deformed star with the rotational frequency $\Omega$ one
has now to deal with a generalized Schwarzschild metric, given by
\cite{48,49}:
\begin{eqnarray}
\label{2.17}
ds^2 = & -\,e^{2\nu(r,\theta,\phi)} dt^2 + e^{2\psi(r,\theta,\Omega)}
   (d\phi - \omega(r,\Omega)dt)^2 + e^{2\mu(r,\theta,\phi)} d\theta^2 
   \nonumber \\
& + e^{2\lambda(r,\theta,\phi)} dr^2 + {\cal O}(\Omega^3) ~.
\end{eqnarray}
Here, $\omega(r)$ denotes the angular velocity of the local inertial
frame, which -- due to the dragging of the local system -- is
proportional to $\Omega$.

The metric functions of Eq.\,(\ref{2.17}) which correspond to
stationary rotation and axial symmetry with respect to the axis of
rotation are expanded up to second order as (independent of $\phi$ and
$t$):
\begin{eqnarray}
\label{2.18}
e^{2\nu(r,\theta,\Omega)} & = & e^{2\phi(r)} \left[ 1 + 2\left(h_0(r,\Omega)
   + h_2(r,\Omega) P_2(\cos \phi)\right)\right] ~,\\
\label{2.19}
e^{2\psi(r,\phi,\Omega)} & = & r^2 \sin^2 \theta \left[ 1 + 2
\left(v_2(r,\Omega) - h_2(r,\Omega)\right) P_2 (\cos\theta)\right] ~, \\
\label{2.20}
e^{2\mu(r,\theta,\Omega)} & = & r^2 \left[ 1 + 2 \left(v_2(r,\Omega) - h_2
(r,\Omega)\right) P_2(\cos\theta) \right] \\
\label{2.21}
e^{2\lambda(r,\theta,\Omega)} & = & e^{2\wedge(r)} \left[ 1 + \frac{2}{r}\,
   \frac{m_0(r,\Omega)G + m_2(r,\Omega) P_2(\cos\theta)}{1 - \gamma (r)}
  \right] ~.
\end{eqnarray}
The angular velocity in the local inertial frame is determined by the
differential equation
\begin{equation}
\label{2.22}
\frac{d}{dr} \left(r^4 a(r) \frac{d\omega}{dr}\right)
  + 4r^3 \frac{da(r)}{dr} \omega(r) = 0 ~, \quad r < R_s ~,
\end{equation}
where $\omega(r)$ is regular for $r = 0$ with $\frac{d\omega}{dr} = 0$.
 $a(r)$ abbreviates
\begin{equation}
\label{2.23}
a(r) \equiv e^{-\phi(r)} \sqrt{1-\gamma(r)} ~.
\end{equation}
Outside the star $\omega(r,\Omega)$ is given by:
\begin{equation}
\label{2.24}
\omega(r,\Omega) = \Omega - \,\frac{2}{r^3} J(\Omega) \quad , \quad
          r > R_s ~.
\end{equation}
The total angular momentum is defined by:
\begin{equation}
\label{2.25}
J(\Omega) = \frac{R_s^4}{6} \left(\frac{d\omega}{dr}\right)_{r=R_s} ~.
\end{equation}
From the last two equations one obtains then an angular frequency
$\Omega$ as a function of central angular velocity $\omega_c =
\omega(r=0)$ (starting value for the iteration):
\begin{equation}
\label{2.26}
\Omega(\omega_c) = \omega(R_s) + \frac{2}{R^3_s} J(\Omega) ~.
\end{equation}
Due to the linearity of Eq.\,(\ref{2.22}) for $\omega(r)$ new values for
$\omega(r)$ emerge simply by rescaling of $\omega_c$. The momentum of
inertia, defined by $I = \frac{J}{\Omega}$, is given by ($a(R_s)=1$):
\begin{equation}
\label{2.27}
I =: \frac{J(\Omega)}{\Omega} = \frac{8\pi}{3} \int^{R_s}_0 dv\,r^4 
   \frac{\epsilon + P}{\sqrt{1-\gamma(r)}} \frac{\omega-\Omega}{\Omega} 
   e^{-\phi} ~.
\end{equation}
Relativistic changes from the Newtonian value are caused by the
dragging of the local systems, i.e. $\bar\omega/\Omega$, the redshift
($e^{-\phi}$), and the space--curvature $\left(
\left(1-\gamma(r)\right)^{-1/2} \right)$. For slowly rotating stars
with low masses, one can neglect the dragging $(\frac{\omega}{\Omega} \to
1)$ and rotational deformations, but we would like to emphasize that
the described treatment is not restricted by low masses and/or slow
rotations.

If one has determined $\omega(r)$, one solves in the next step the
coupled mass monopole equations $(\ell = 0)$ for $m_0, p_0$ ( =
monopole pressure perturbation) and $h_0$ (for details, see
Refs.\,\cite{4,7,46}). The quadrupole distortions $h_2$ and $v_2$
$(\ell = 2)$ determine the star's shape (see
Refs.\,\cite{4,46}). After the determination of the distortion
functions, one can express the surfaces of constant density, the
star's eccentricity \cite{50}, and the mass quadrupole moment of the
star as follows:
\begin{eqnarray}
\label{2.28}
r(\theta) & = & r + \xi_0(r) + \left\{\xi_2(r) + r[v_2(r) - h_2(r)]\right\}
   P_2(\cos\theta) ~, \\
\label{2.29}
e &  =  &  \sqrt{1-\left(\frac{R_p}{R_{eq}}\right)^2} , \\
\label{2.30}
\Pi &  =  & \frac{8}{5}~A_2 \left(\frac{\gamma_s}{2} \right)^2 + 
\left(\frac{J}{R_s} \right)^2 ~.
\end{eqnarray}
$A_2$ denotes an integration constant and $\xi_i$ is defined by
($i=0,2$):
\begin{equation}
\label{2.31}
\xi_i = -\,p_i (\epsilon + P) \left(\frac{\partial P}{\partial r}
\right)^{-1} ~.
\end{equation}

\vskip2ex
\noindent
{\bf c) Stability criteria}

\vskip1ex
An intriguing problem in the physics of NSs is the question, whether
the used EOSs are in accordance with the observed data, so supplying a
test for the theoretical and partly speculative EOSs for highly
compressed matter.  Unfortunately, as far as gross properties (radii,
masses) of NSs are concerned most of the nonrelativistic and
relativistic EOSs are able to reproduce these gross properties (For a
test of 25 different EOSs, see Refs.\,\cite{4,7,24}).

A more decisive criterion may be the stability of a star against
rotation.  Since no trivial stability criteria are known for rotating
configurations in general relativity, we consider first Kepler's
criterion, which sets an absolute upper limit on a star's
rotation. The resulting Kepler frequency, $\Omega_K$, beyond which
instability sets in due to mass shedding at the equator, is given for
the generalized Schwarzschild metric as solution of ($\psi' \equiv
\frac{\partial \psi}{\partial r}$, etc.) \cite{4,7,46,47}
\begin{equation}
\label{2.32}
\Omega = \left[ e^{\nu(\Omega)-\psi(\Omega)} V(\Omega) +
\omega(\Omega)\right]_{eq, \Omega=\Omega_K} ~,
\end{equation}
with
\begin{equation}
\label{2.33}
V(\Omega) := \left[ \frac{\omega(\Omega)'}{2\psi(\Omega)'}
e^{\psi(\Omega)-\nu(\Omega)}\right] + \left.\sqrt{\frac{\nu(\Omega)'}{\psi(\omega)'}
 +   \left( \frac{\omega(\Omega)'}{2\psi(\Omega)'}
e^{\psi(\Omega)-\nu(\Omega)}\right)^2} \right|_{eq,\Omega=\Omega_K} ~,
\end{equation}
to be evaluated at the equator. $V$ denotes the orbital velocity of a
comoving observer at the equator relative to a locally non--rotating
observer. Neglection of the distortions $(h_2,v_2<<1)$ and of the dragging of
the local inertial frames $(\omega = \Omega)$ gives
\begin{equation}
\label{2.34}
V_{eq} = \sqrt{\frac{\gamma_{eq}}{2}} \frac{1}{\sqrt{1 - \gamma_{eq}}} \to
R_{eq} \Omega_c  \quad \mbox{or} \quad \gamma_{eq} \to 1 ~,
\end{equation}
with 
\begin{equation}
\label{2.35}
\Omega_c \equiv  \sqrt{\frac{M_s}{R^3_s}} ~.
\end{equation}
The Newtonian result is recovered by using a flat
space--time--geometry. The Kepler frequency of the heaviest neutron
star can be obtained from the mass and radius of the most massive
nonrotating neutron star as
\begin{equation}
\label{2.36}
\Omega_K = \frac{2}{3} \sqrt{\frac{M_s}{R_s^3}} ~,
\end{equation}
which has the advantage that one needs only the input from a static
non--rotating star model.

As mentioned before the Kepler criterion gives only an upper limit.
Gravitational--wave reaction instabilities of a rotating star are
likely to lower the star's maximum rotational frequency below
$\Omega_K$. Since the theory is rather lengthy, we refer for details to
Refs.\,\cite{4,7,46,47,51,52}. The critical frequency for a particular
instability mode $(m = 2,3 \ldots)$ is given by:
\begin{equation}
\label{2.37}
\Omega^\nu_m = \frac{\omega_m(0)}{m} \left\{ a_m (\Omega^\nu_m) +
  \gamma_m(\Omega^\nu_m) \left(\frac{\tau_{g,m}}{\tau_{r,m}}
\right)^{\frac{1}{2m+1}}\right\}
\end{equation}
where $\nu$ denotes the shear viscosity, depending on temperature $T$. The
expressions for the damping time scales $\tau$ for gravitational radiation
reactions $(\tau_{g,m})$, viscous damping $(\tau_{v,m})$, and the surface
mode $\omega_m$ can be found in Refs.\,\cite{4,7,46,51,52}.

\vskip2ex
\noindent
{\bf d) Further quantities}

\vskip1ex
For completeness we also give the expressions for the redshifts, the
injection energy, and the stability parameter.

The frequency shifts of light emitted at the equator in backward (b) and
forward direction (f) is given by \cite{4,7,46}:
\begin{equation}
\label{2.38}
z_{\rm b/f} (\Omega) = e^{-\nu(\Omega)} \left( 1\pm
\omega(\Omega)\,e^{\psi(\Omega)-\nu(\Omega)} \right)^{-1} 
  \left( \frac{1\pm V(\Omega)}{1\mp V(\Omega)} \right)^{1/2} -1 ~.
\end{equation}
For the redshift at the pole one gets:
\begin{equation}
\label{2.39}
z_p(\Omega) = e^{-\nu(\Omega)} -1 ~.
\end{equation}
The so--called injection energy is defined as:
\begin{equation}
\label{2.40}
\beta(\Omega) \equiv \left. e^{2\nu(\Omega)}\right|_{\rm pole} =
    \frac{1}{(z_p(\Omega) + 1)^2} ~.
\end{equation}
For discussing the stability of rotating stars it is useful to define the
stability parameter:
\begin{equation}
\label{2.41}
t(\Omega) := \frac{T(\Omega)}{|W(\Omega)|} ~,
\end{equation}
where $T$ denotes the rotational and $W$ the gravitational energy of the
star.

\vskip2ex
\noindent
{\bf e) Results (gross properties)}

\vskip1ex
As a first example for the influence of the rotation we show in
Figs.\,13 and 14 the NS--mass versus central energy density and the
radius--mass relation for a composition with $p,n,e^-$, and $\mu^-$
only, for nonrotating stars and stars rotating with their Kepler
frequency. As expected and discussed before one reaches, due to the
large stiffness of the EOS, the maximum mass at lower central
densities. The rotation can increase the mass by more than half a
solar mass. In this context one may remark that relativistic EOS of
nuclear matter are in general stiffer than their nonrelativistic
counterparts
\cite{7,24,28}. If one includes now hyperons in the relativistic NSM--EOSs
one obtains a softening of the EOS, so that the gross properties of NSs may
not differ too much from NS--calculations with nonrelativistic
treatments with a pure nucleonic/leptonic composition. Only the protons
concentration are not the same in both cases, since the different behaviour
of the symmetry energy lowers the proton contribution in the nonrelativistic
case, so suppressing the so--called direct Urca--process \cite{53,54,55}.

Next we turn to the more interesting case of NSM including more
baryons. For EOSs in the relativistic Hartree scheme one obtains -- as
expected -- still sufficiently large enough NS--masses, which decrease
-- as discussed before -- for weaker hyperon couplings. This behaviour
is exhibited in Figs.\,15 and 16. So far the results comply with the
familiar pattern. A limit for the relative hyperon couplings, in this
scheme, is given by the SU(6) choice, since otherwise the NS--masses
become too small. This result compares with our NS--calculation with
the NSM--EOS TM\,1, where we obtained a maximum star mass of
1.56\,M$_\odot$ (see Section\,II). Furthermore we find the expected
peculiar behaviour for the universal coupling in the lower
energy--density region (see discussion in Section\,II). This can also
be seen in Table\,V, where the NS--properties for a fixed mass of
1.4~M$_\odot$ are given. They show with exception of the universal
coupling the expected behaviour with increasing
hyperon--couplings. For comparison we include also the outcome for
nonrelativistic calculations without hyperons \cite{23}, which
demonstrates clearly the softening of the relativistic EOS due to the
hyperons (see also Figs.\,13,15,16), which become even softer than the
nonrelativistic EOS without hyperons.

More interesting are the RHF--EOSs. In the case of using the hyperon
couplings of the RH--treatment and $g_N=g_\Delta$ one obtains for
rather low densities an onset of $\Delta^-$ and hyperons, where the
$\Delta^-$ and the $\Sigma^-$ play a decisive role (neglecting the
$\Delta^-$ does not change the EOS significantly, since then the
$\Sigma^-$ occur earlier \cite{18}), and the EOS becomes rather soft
for lower densities, but for higher densities the pressure rises again
more strongly. As a consequence the gravitational NS--mass as a
function of central energy density should show a flat plateau before
it rises again in the standard pattern, but at higher densities than
in the RH--EOS case.  Furthermore one expects relatively low star
masses due to the soft EOS.
 
This behaviour is exhibited in Fig. 17, where the gravitational
NS--mass is given as function of the central density. Reasonable star
masses demand larger hyperon couplings.  The flat behaviour in the
lower domain resembles -- as one could have expected -- to the case of
rather soft EOSs with low incompressibility (see, for instance, the
BCK--EOS \cite{42}, where one obtains maximal NS--masses of
approcimately 1~M$_\odot$ \cite{41}).  In Fig.\,18 we exhibit the
limiting Kepler frequencies. As expected the softer RHF--EOSs permit
lower rotational periods. However inclusion of the
gravitational--radiation instabilities shows that the Kepler frequency
is only an upper limit on the critical frequency and the limiting
periods increase of about 30\,\% \cite{4,7,10} (for more details, see
Ref.\,\cite{18}). If one uses RHF--EOSs, where the ratio of hyperon
couplings is adjusted in the RHF--scheme (see Section\,II), the
hyperons are not so easily produced and the resulting EOSs become
stiffer (see Section\,II). In the presentation we will restrict
ourselves to the interesting cases of the smallest hyperon couplings
compatible with star masses around $1.5\msun$.  The resulting star
masses as function of the central energy density are shown in
Figs.\,19 and 20. By comparison with Fig.\,17 we infer that these
improved EOSs, where the hyperon potential depths are treated
correctly, lead to masses, which are better in accordance with the
data, since increase of the hyperon couplings and rotation lead to
even higher star masses. EOSs with weak $\Delta$--repulsion are rather
soft and give only for strong hyperon couplings and high rotation
frequencies mass values around 1.5~$\msun$ (see Figs.\,19,20).

One may illuminate the situation further by comparing the properties
for a typical NS of 1.4 $M_\odot$ (see Tabs. V,VI). By the given
arguments one should obtain increasing central densities by going from
the stiff RH-EOS (RH1) to the softest RHF-EOSs with low Dirac masses
for both the deltas and hyperons (RHF12,13). For the properly adjusted
RHF-EOSs (RHF8,9) the star parameters differ not significantly. Here
the higher $\Delta$-masses suppress the influence of the $\Delta$'s
below $\epsilon\sim 500{\rm ~MeV/fm}^3$ and one obtains sufficient
pressure to reach a mass of 1.4 $M_\odot$ earlier (see also
Fig. 19). Decrease of the $\Delta$-mass according to universal
coupling of the $\Delta$'s gives an increase of the central energy by
a factor 2 (RHF1 compared with RH1). Finally we illustrate in Table
VII the influence of rotation. Shown are the results for a NS with the
same baryon number. The mass changes due to rotation are relatively
small for constant baryon number (for fixed central energy density see
Refs [4,7,24]).

\vskip2ex
\noindent
{\bf f)  Cooling properties}
\vskip1ex

Another decisive test for the EOS may be the cooling history of a
NS. In Ref. \cite{55} we have already calculated and discussed the
thermal evolution of various models with different EOSs and different
involved processes. Here we show and compare the cooling tracks of NS
models constructed for the three EOSs RH1, RHF1, and RHF8, as well as
the RHF-EOS without hyperons (see Fig. 22). During about the first
million years the NS cools mainly by emission of neutrinos. One
classifies the neutrino processes into slow and enhanced ones,
according to whether two or only one baryon is participating. Enhanced
processes cause a temperature inversion in young stars, i.e. the
interior of the star becomes much cooler than the crust. Depending on
the crust thickness the cooling wave formed by the temperature
gradient reaches the surface and causes the sharp decrease of the
surface temperature after about 30 years (see the three broken curves
in Fig. 22).

For the EOSs considered here the only possible enhanced cooling
processes are the nucleon \cite{36,37}
\begin{equation}
  {\rm n} \rightarrow {\rm p}+l^- +\bar\nu_l
\end{equation}
and the hyperon direct Urca processes \cite{Prakash92}
\begin{equation} \label{eq:dirurca1}
  \Sigma^- \rightarrow \Lambda+l^- +\bar\nu_l
\end{equation}
and
\begin{equation}
  \Lambda \rightarrow {\rm p}+l^- +\bar\nu_l~,
\end{equation}
as well as their inverse reactions. Here, $l$ denotes electrons and
muons. The nucleon direct Urca process is only possible, if the proton
fraction exceeds some critical value of about 11 \% for a pure
nucleonic/electron composition of the neutron star matter, since
otherwise energy and momentum conservation cannot be fulfilled
simultanously. If hyperons and muons are taken into account this value
rises slightly to approximately 13 \%. Similar constraints have to be
considered for the hyperon direct Urca processes. It is obvious that
the resulting thermal evolution depends strongly on the EOS. It seems
to be a general feature that non-relativistic EOSs have protron
fractions below this critical value \cite{25}, whereas relativistic
EOSs allow for the nucleon direct Urca process. The critical masses
above which the hyperon direct Urca processes are possible are
approximately equal to 1.3 $M_\odot$ for all three EOSs studied in
this section. This seems to be surprising, since $\Lambda$ and
$\Sigma^-$ appear beyond $n\sim 0.7{\rm ~fm}^{-3}$ in the case of RHF1
(see Fig. 9), and already beyond $n\sim 0.3{\rm ~fm}^{-3}$ in the cases
of RH1 and RHF8 (see Figs. 6 and 11). This higher threshold density is
however compensated by the smaller incompressibility of RHF1 compared
to RH1 and RHF8 (see Sect. II). The used slow neutrino processes, as
well as the processes in the crust of the NS, are discussed in greater
detail in Ref. \cite{56}.

The cooling behaviour of a NS is also influenced by the appearance of
superfluid phases. If neutrons or protons become superfluid the
neutrino emissivity of the nucleonic processes, the thermal
conductivity and the heat capacity are reduced by an approximately
exponential factor $\exp(-\Delta/kT)$, where $\Delta$ denotes the gap
energy (see Table IV of Ref. \cite{56} for the used gap energies).

The observational data are described in Ref. \cite{Schaab97b} (see
Table 2 in Ref. \cite{Schaab97b}). The obtained effective surface
temperature depends crucially on whether a magnetized hydrogen
atmosphere is used or not. Since the photon flux, measured solely in
the X-ray energy band, does not allow to determine what atmosphere one
should use, we consider both the blackbody model (solid error bars in
Fig. 22) and the hydrogen-atmosphere model (dashed error bars). The
plotted errors represent the $3\sigma$ error range due to the small
photon fluxes.

All models exhibit enhanced cooling via the nucleon direct Urca
process. However this process is suppressed below the critical
temperature for the superfluid phase transition. Since the process
(\ref{eq:dirurca1}) is not suppressed by superfluidity, the surface
temperature of these models (see broken curves) is much smaller than
the one of the model without hyperons (solid curve). The observed
data can almost perfectly be described by the latter model, provided
one assumes that the pulsars have no hydrogen atmosphere (except PSR
1055-52, which could be explained by internal heating; see,
e.g. Refs. \cite{VanRiper95a,Schaab97c}). However, if some of the
pulsars prove to have a hydrogen atmosphere, these models seem to be
too hot. Whether the observed pulsars have a hydrogen atmosphere could
be decided if one considers multiwavelength observations, as suggested
by Pavlov et al. \cite{Pavlov96a}.  With respect to cooling properties
one seems to get along with simpler relativistic EOSs without
hyperons. However the weaker temperature drop in this case is caused
by the superfluidity of nucleons, which cannot be included, at
present, for the other baryons. Inclusion of this effect for hyperons
may shift the curves towards the observed values, since the process
(\ref{eq:dirurca1}) would be suppressed, too.

Please note that we have considered only some of the possibilities of
neutron star cooling. Additional processes as internal heating
\cite{VanRiper95a,Schaab97c} or intermediate neutrino processes
\cite{Schaab95b} may yield different cooling tracks. This is also true
for the effect of accreted atmospheres investigated in
Refs. \cite{Potekhin96c,Page96b}.

\section{Summary}

The goal of this investigation is to incorporate ``parameter--free''
microscopic relativistic Brueckner--Hartree--Fock calculations of
nuclear matter in the investigation of neutron star matter. In a first
step we extended the RBHF--theory of asymmetric nuclear matter to
neutron star matter, consisting of neutrons, protons and leptons,
which has to obey the constraints of charge neutrality and generalized
$\beta$--equilibrium. Since for higher densities more baryons
(hyperons etc.) have to be included, for which, at present,
microscopic RBHF--calculations are not feasible, we extended the
scheme by utilizing either the relativistic Hartree-- or
Hartree--Fock--approximation, in which the other baryons can be
incorporated.  The coupling constants of these schemes were adjusted
in the nucleonic sector to the outcome of the RBHF--calculations of
NSM around saturation densities.  In this manner we obtained a good
description of NSM near saturation, which is essential for lighter
neutron stars and also important for heavier stars.  As long as one
restricts the composition to $p,n,e^-$, and $\mu^-$ only, one obtains
in this framework, due to the rather stiff EOS, Oppenheimer--Volkoff
star masses around 2.2~M$_\odot$ and minimum rotation frequencies
slightly above 1~ms. However if one incorporates more baryons in the
scheme one obtains a considerable softening of the NSM--EOS, which
leads inevitably, as in the case of realistic phenomenological
parametrizations of the nuclear Lagrangian, to negative Dirac masses
for the nucleons in NSM. This drawback is the result of the necessity
to reproduce a rather low nucleon Dirac mass at saturation, which
leads to a peculiar feature of the mean field approximation, namely
that such Lagrangians cause a strong decrease of the effective
$\sigma$--mass at higher densities, resulting in a steep decrease of
the Dirac masses. Since the behaviour of the Dirac masses at high
densities are (completely) unknown in NSM, we extrapolated the
Lagrangian in these domains by a slightly changed dynamics, where the
decrease of the Dirac mass is not so severe, which is also in
accordance with RBHF--results in nuclear matter. The resulting
pressure changes are rather small. Within the RH--framework we
obtained then compositions of NSM which are more or less in accordance
with former investigations, which use a priori in the whole domain
Lagrangians with increasing effective $\sigma$--masses. For the
hyperons we used different coupling strengths but the ratios of the
$\sigma-\omega$--couplings were fixed by utilizing the potential
depths of the hyperons in nuclear matter. Due to large
$\rho$--coupling in the RH--approximation the $\Delta$'s are
negligible in this framework. The Oppenheimer--Volkoff (OV) star mass
reaches from 1.5~M$_\odot$ -- 2.4~M$_\odot$ depending on the hyperon
couplings and the rotation frequencies. More complicated is the
situation for RHF--EOSs. Here one is confronted with smaller
$\rho$--couplings, which favours $\Delta$'s in the composition, and
the fact that the potential depths of the hyperons in nuclear matter
are solely determined by their Hartree contributions. As discussed in
details in Section\,II, the resulting compositions depend now strongly
on both the assumptions about the hyperon couplings and the assumed
behaviour of the $\Delta$--Dirac masses in NSM. As long as one assumes
for the $\Delta$'s the same coupling as for the nucleons, the
$\Delta$'s play a dominant role in the composition. The EOSs are
relatively soft and the minimal Oppenheimer--Volkoff mass is around
1.5~M$_\odot$ for weak hyperon couplings. This mass drops even to
1.3~M$_\odot$ for strong hyperon couplings if one decreases the
relative $\Delta$--strength generally to 0.625\,. Increase of the
$\Delta$--repulsion gives compositions not so different from the
RH--approximations with minimum OV--masses of 1.6~M$_\odot$.

In general we can conclude in accordance with earlier findings that the
EOSs based on hadronic theories of matter are capable of accomodating
the gross properties as well as rotational periods of all pulsars
known to date.

\bigskip
\noindent
{\bf Acknowledgements}

We would like to thank M.~Mare\u s, J.~Schaffner, W.~Wambach and W.~Weise for
helpful comments and clarifying remarks.


\newpage
\section*{Table captions}
\begin{description}
\item[Table\,I:] Saturation properties of infinite nuclear matter. RBHF--$A$
and RBHF--$B$ denote the RBHF--results for the Brockmann potentials $A$ and
$B$, calculated in the full Dirac space (for details, see Ref.\,\cite{21}).
For
comparison we added some results of more phenomenological relativistic
Hartree calculations (NL1, NL--SH) \cite{23} and phenomenological
nonrelativistic calculations. SkM$^*$ and S\,III denote two well known Skyrme
forces, TF\,96 is a recent Thomas--Fermi calculation \cite{24}. Also
included
are two nonrelativistic microscopic variational calculations \cite{25}.
\item[Table\,II:] Parametrizations of the RH-- and RHF--Lagrangian adjusted
to the RBHF--calculations.  For the masses the following values were selected
(MeV):
$m_N = 939$, $m_\sigma = 550$, $m_\omega = 738$, $m_\pi = 138$, $m_\rho =
770$ ($g_\pi = 1.00265~f^2_\pi/4\pi = 0.08$). The parametrizations are labelled
as follows: RHA
$\hat =$ RBHA etc. for the potential $A$ (effective mass at the Fermi
surface $\tilde m = 617.8$~MeV\,($A$); 621.8~MeV~($B$)).
\item[Table\,III:] EOSs for the different density regions of a NS
(1~MeV/fm$^3$ corresponds to 1.783 $\times 10^{12}$~g/cm$^3$). For the
density region above 20~MeV/fm$^3$ we use the parametrizations of the
RBHF--calculations in the frame of the RH-- and RHF--approximation,
respectively (OBE--potential B of Brockmann and Machleidt). The parameters
for the nucleonic sector are given in Table\,II. For the population of the
more massive baryons the 12 lowest lying ones are allowed for.
\item[Table\,IV:] Relative coupling strengths of the hyperons in the
different approximations (see text).  The ratios $x_{\sigma H}/x_{\omega H}$
are adjusted to the binding energy of the hyperon in nuclear matter. For
RHF11 - RHF14 the RH--couplings are used. The universal coupling is defined
by $g_N = g_H = g_\Delta$. The EOSs for RHF7, RHF11 and RHF12 are too soft
for obtaining OV--masses around 1.5~$\msun$ (see Fig.\,17). In these models
larger hyperon couplings are needed, for instance, by going from RHF7 to
RHF10 M$_{OV}$ increases approximately to 1.5~$\msun$ (at Kepler frequency).
\item[Table\,V:] Comparison of the properties of a static, spherical NS of
mass 1.4~M$_\odot$ for different RH--models. $\epsilon_c(P_c)$ denotes the
central energy density (pressure). The $amu$ mass $M_A$ minus the
gravitational mass $M_G$ is effectively the binding energy liberated when the
NS is formed. $R$ denotes the star's radius, $\Delta_c$ stands for the
stellar crust using 2.4 $\times 10^{14}$~g~cm$^{-3}$ as the boundary, $I$
denotes the moment of inertia and $z$ the surface redshift. For a comparison
we included two nonrelativistic models without hyperons (see Table\,I)
\cite{23}.
\item[Table\,VI:] 
Comparison of the properties of a static, spherical NS of mass
$M=1.4M_\odot$ for different RHF models. Labels as in Table V.
\item[Table\,VII:] 
Properties of rotating neutron star models of rotational period
$P=1.4$~ms and the same baryon number as the nonrotating star with
$M=1.4M_\odot$, calculated for different EOSs. The entries are:
central energy density $\epsilon_{\rm c}$; equatorial and polar radii,
$R_{\rm eq}$ and $R_{\rm p}$, respectively; moment of inertia $I$;
stability parameter, $t$; injection energy $\beta$; redshift of the
pole, $z_{\rm p}$; eccentricity, $e$ \cite{49}; quadrupole moment,
$\Pi$. The gravitational mass increase due to rotation is rather
small.
\end{description}

\newpage
\section*{Figure captions}
\begin{description}
\item[Fig.\,1:] EOSs for asymmetric nuclear matter. Compared are the EOSs 
for the Brockmann--Machleidt potential B for a fixed asymmetry
$\delta$ in the RBHF--approximation (RBHF--B: full curves) with the
treatment in the RH (RHB: dotted curves) and RHF (RHFB\,1: dashed
curves) --scheme, respectively.
\item[Fig.\,2:] Comparison of the EOSs for neutron star matter composed of
$n,p,e^-$, and $\mu^-$ (RBHF-B, RHB, RHFB\,1; Brockmann potential B). The
upper branches correspond to the case where myons are neglected.
\item[Fig.\,3:] Comparison of the nucleon/lepton composition of neutron star
matter (potential B). 
The upper branches correspond to the case without myons.
\item[Fig.\,4:] Comparison of the Dirac masses with zero charge: 
The full curves show the behaviour in the original scheme. The dotted
curves give the extrapolation according to
Eqs.(\ref{II.10n},\ref{II.11n}). The selected case is the
RHF--approach with SU(6) couplings (RHF11, $\alpha=0.02$) (see
Table\,IV).  The other cases are rather similiar. The masses differ
hardly for the different isospin states.
\item[Fig.\,5:] Relative baryon/lepton populations in the relativistic
Hartree scheme for universal coupling of the hyperons and deltas (RH1;
$\alpha = 0.015$)
\item[Fig.\,6:] Relative baryon/lepton populations in the relativistic
Hartree scheme  for hyperon--SU(6)--coupling of the vector mesons (RH1;
$\alpha = 0.015$). The relative $\sigma$--couplings of the hyperons are
adjusted to the corresponding potential depths in nuclear matter. 
\item[Fig.\,7:] Dependence of the EOS on the choice of the relative
$\sigma$--meson--hyperon coupling $x_{\sigma H}$ (relativistic Hartree EOS,
potential B, $\alpha = 0.015$). Compared are the choices $x_{\sigma H}= 0.7,
0.8$ with the so--called SU(6) parametrization and the universal coupling
(see Table\,IV). For the first two cases the (larger) hyperon-vector
couplings are adjusted to the potential depths in infinite matter (for
$x_{\sigma H}= 0.9, x_{\omega H}>1$). For the SU(6) parametrization the
vector meson couplings are fixed by the quark picture and the (smaller)
$\sigma$--couplings are adjusted to the potential depths ($x_{\sigma H}\leq
x_{\omega H}$). Due to the stronger $\sigma$--coupling in the universal case
one gets for lower densities a softer EOS, otherwise one confirms the
expectation that smaller couplings softens the EOS (see text).
\item[Fig.\,8:] Relative baryon/lepton populations in the RHF--approximation
 for relative hyperon couplings taken from the RH--approximation in the
SU(6) case (RHF11; $\alpha = 0.02$)
\item[Fig.\,9:] Relative baryon/lepton population in the RHF--approximation
(RHF1; $\alpha = 0.02$). The $\sigma$--couplings are adjusted to the hyperon
potential depths (see Table\,IV). Increase of the hyperon coupling would
give a later onset of the hyperons, for instance, $\Lambda$ occurs at $\rho
\sim 0.8$~fm$^{-3}$.
\item[Fig.\,10:] Relative baryon/lepton population for NSM in the
RHF--approximation with reduced $\Delta$--coupling ($x_{\omega H}$ = SU(6),
$x_{\sigma H}$ adjusted, (RHF7; $\alpha = 0.02$), $x_{\Delta\Delta} =
x_{\Delta N} = 0.625$ \cite{44}). The onset of the $\Delta$s/hyperons is now
shifted to higher/lower densities than for $g_\Delta = g_N$.
\item[Fig.\,11:] Relative baryon/lepton population for NSM in the
RHF--approximation (RHF8) with fixed $x_{\sigma\Delta} = 0.625$ and
$x_{\omega\Delta} = 1$. The hyperon couplings are adjusted to the
hyperon--potential depths.
\item[Fig.\,12:] Relative baryon/lepton population for NSM in the
RHF--approximation (RHF9) with adjusted delta-- and hyperon couplings to the
potential depths.
\item[Fig.\,13:] Gravitational star mass (in units of solar Mass M$_\odot$)
as a function of central energy density for star models constructed from EOSs
 with $p,n,e^-$, and $\mu^-$. The upper curve corresponds to (deformed)
stars rotating at their Kepler frequency. The RHF-- and RH--curves are very
close, since the pressure differs not much.
\item[Fig.\,14:] Neutron star radius versus mass. Shown are sequences of
stars rotating at their Kepler frequencies and at zero frequency ($p,n,e^-$,
and $\mu^-$ only). The minimal periods are $\sim$ 1~ms for $M=1.5M_\odot$.
\item[Fig.\,15:] Dependence of the gravitational star mass on the relative
meson--hyperon coupling. The non rotating, spherical star families are given
as function of the central energy density (RH--EOSs, $\alpha = 0.015$; all
baryons included). The maximum star mass increases with stronger hyperon
couplings. The peculiar behaviour for the universal coupling for lower
densities is explainable by the stronger attraction (see Figs.\,5,7 and
text).
\item[Fig.\,16:] Increase of the gravitational star mass for rotating stars:
Shown are the star families as in Fig.\,15, but now for deformed stars
rotating at their Kepler frequency. The radii are approximately 12 (15)~km
for nonrotating (rotating) stars at $M_{NS}=1.5~M_\odot$.
\item[Fig.\,17:] Gravitational NS mass for nonrotating stars versus central
energy density for RHF--EOSs with universal couplings and couplings from the
RH--treatment. The peculiar behaviour for low densities corresponds to the
early onset of baryons in this model (see text). At their Kepler frequency
the OV--mass ranges from $\sim 1.95~M_\odot$ (universal) to
$\sim 1.5~M_\odot$ (SU(6)--model); for $M = 1.4 M_\odot$ the radius
stretches in the range 9~km (nonrotating) to 11~km (Kepler rotating).
\item[Fig.\,18:] Limiting rotational Kepler periods of pulsars versus NS mass
for the models described in Figs.\,16 and 17. The shaded area covers the
range of observed periods and masses \cite{56}.
\item[Fig.\,19:] 
Gravitational NS mass for spherical nonrotating stars versus central energy
density. Compared are the star families for EOSs in the RHF--scheme for
different $\Delta$--couplings for the weakest hyperon coupling compatible
with $M_{OV}\sim 1.5~\msun$. The ratio of the hyperon couplings is adjusted
to the hyperon potential depths in nuclear matter.
\item[Fig.\,20:] Gravitational NS mass for nonspherical stars rotating at
their Kepler frequency versus central energy density. Compared are the same
EOSs as in Fig.\,19.
\item[Fig.\,21:] Limiting rotational Kepler periods of pulsars versus NS mass
for the models described in Figs.\,19, 20. Inclusion of gravitation--radiation
instabilities increases the limiting period by approximately 30\% (for $M =
1.5~\msun$) \cite{18}. The shaded area covers the range of observed periods
and masses \cite{56}.
\item[Fig.\,22:]
Cooling of $M=1.4M_\odot$ models for different EOSs. The surface
temperature obtained with a blackbody- (magnetic hydrogen-) atmosphere
are marked with solid (dashed) error bars labeld by the respective
pulsar's position.
\end{description}
\clearpage\newpage
\renewcommand{\baselinestretch}{1.3} \small\normalsize
\begin{table}
\begin{center}
TABLE I \\[1cm]
\begin{tabular}
{|l|l|c|c|c|}
\hline  \hline
 & ~E/A & $\rho_{00}$  &  $K_v$  &  $J$  \\
 & (MeV) & (fm$^{-3}$) & (MeV)   & (MeV) \\
\hline
RBHF--A &  -16.49 & 0.174 & 280 & 34.4  \\
RBHF--B & -15.73 & 0.172 & 249 & 32.8 \\
NL1  & -16.4 & 0.152 & 212 & 43.5 \\
NL--SH & -16.3 & 0.146  & 356 & 36.1 \\
SkM$^*$ & -15.8 & 0.160 & 216 & 30.0 \\
S\,III & -15.9 & 0.145 & 355 & 28.2 \\
TF96 & -16.24 & 0.161 & 234 & 33 \\
WUU & -15.5 & 0.175 & 202 & 30 \\
WUT & -16.6 & 0.157 & 261 & 29 \\
\hline
\end{tabular}

\end{center}

\end{table}

\newpage
\begin{table}
\begin{center}
TABLE II \\[1cm]
\begin{tabular}
{|l|r|r|r|r|r|c|}
\hline
 & $g_\sigma$~~ & $g_\omega$~~ & $g_\rho$~~ & $10^3 \times b_N$ & $10^3
\times c_N$ & $f_\rho/g_\rho$ \\
\hline
RHA & 9.58096 & 10.67698 & 3.81003 & 3.333665 & -3.52365 & -- \\
RHF\,A1  & 9.28353 & 8.37378 & 2.10082 & 3.333689 & -2.15239 & -- \\
RHF\,A2 & 9.24268 & 8.25548 & 2.19809 & 2.96514 & -2.68614 & 3.7 \\
RHF\,A3 & 9.16665 & 8.07540 & 2.37987 & 1.95524 & -2.36335 & 6.6 \\
RHB   &  9.59169 & 10.68084 & 3.66541 & 3.62616 & -4.17140 & -- \\
RHF\,B1 & 9.36839 & 8.40466 & 1.77326 & 3.74354 & -3.18456 & -- \\
RHF\,B2 & 9.33266 & 8.32154 & 1.86078 & 3.44306 & -3.46261 & 3.7 \\
RHF\,B3 & 9.26782 & 8.19391 & 2.02216 & 2.67292 & -3.18198 & 6.6 \\
\hline
\end{tabular}
\end{center}
\end{table}

\begin{table}
\begin{center}
TABLE III \\[1cm]
\begin{tabular}{|lcl|}
\hline
EOS & Energy--density range & composition \\
 &   (MeV~fm$^3$)  & \\
\hline
HW$^a$ & $\epsilon < 0.6$  &  Crystalline; light metals, electron gas \\
NV$^b$ & $0.6 < \epsilon < 20$ & Metals, relativistic electron gas \\
RHB  &  $20 < \epsilon$ & $ p,n,e^-,\mu^-$ \\
RHF\,B1  & $20 < \epsilon$ & $p,n,e^-,\mu^-$ \\
RHB  & $20 < \epsilon$ & $p,n,\Lambda,\Sigma^{\pm,0},
                           \Xi^{0,-},\Delta{\rm 's},e^-,\mu^-$ \\
RHF\,B1  & $20 < \epsilon$ & $p,n,\Lambda,\Sigma^{\pm,0}
                              \Xi^{0,-},\Delta{\rm 's}, e^-,\mu^-$ \\
\hline
\end{tabular}
\end{center}
$^a$ taken from Ref.\,\cite{19}, \\
$^b$ taken from Ref.\,\cite{20}
\end{table}

\begin{table}
\begin{center}
TABLE IV \\[1cm]
\begin{tabular}{|l|l|l|l|l|l|l|}
\hline
Approximation & $x_{\sigma\Sigma\Lambda}$ & $x_{\omega\Sigma\Lambda}$ &
      $x_{\sigma\Xi}$ & $x_{\omega\Xi}$ & $x_{\sigma\Delta}$ &
$x_{\omega\Delta}$\\
\hline
\hline
RH1(SU(6)) & 0.676 & 2/3 & 0.342 & 1/3 & 1 & 1 \\
\hline
RH2 & 0.7 &  0.77 & 0.7 & 0.78 & 1 & 1 \\
\hline
RH3 & 0.8 & 0.9 & 0.8 & 0.91 & 1 & 1 \\
\hline
RH4 & 0.9 & 1.03 & 0.9 & 1.04 & 1 & 1 \\
\hline
RH5(U)  & 1 & 1 & 1 & 1 & 1 & 1 \\
\hline
RHF1(SU(6)) & 0.44 & 2/3 & 0.26 & 1/3 & 1 & 1 \\
\hline
RHF2 & 0.36 & 0.5 & 0.35 & 0.5 & 1 & 1 \\
\hline		         	
RHF3 & 0.46 & 0.7 & 0.45 & 0.7 & 1 & 1 \\
\hline		         	
RHF4 & 0.51 & 0.8 & 0.5  & 0.8 & 1 & 1 \\
\hline		         	
RHF5 & 0.56 & 0.9 & 0.55 & 0.9 & 1 & 1 \\
\hline
RHF6(U) & 1 & 1 & 1 & 1 & 1 & 1 \\
\hline
RHF7(SU(6)) & 0.44 & 2/3 & 0.26 & 1/3 & 0.625 & 0.625 \\
\hline
RHF8(SU(6)) & 0.44 & 2/3 & 0.26 & 1/3 & 0.625 & 1 \\
\hline
RHF9(SU(6)) & 0.44 & 1/3 & 0.26 & 1/3 & 0.75 & 1 \\
\hline
RHF10  & 0.51 & 0.8 & 0.5 & 0.8 & 0.625 & 0.625 \\
\hline
RHF11(SU(6))  & 0.676 & 2/3 & 0.342 & 1/3 & 1 & 1 \\
\hline
RHF12  & 0.7 & 0.77 & 0.7 & 0.78 & 1 & 1 \\
\hline
RHF13  & 0.8 & 0.9 & 0.8 & 0.91 & 1 & 1 \\
\hline
RHF14  & 0.9 & 1.03 & 0.9 & 1.04 & 1 & 1 \\
\hline
\end{tabular}
\end{center}
\end{table}

\begin{table}
\begin{center}
TABLE V \\[1cm]
\begin{tabular}{|l|c|c|c|c|c|c|}
\hline
Quantity & RH1(SU(6)) & RH2 & RH3 & RH5(U) & WUT & TF \\
\hline
\hline
$\epsilon_c~(10^{14}$g/cm$^3$) & 8.1 & 7.7 & 7.2 & 8.9 & 12.12 & 10.30 \\
\hline
P$_c~(10^{34}$dyn/cm$^2$) & 8.98 & 8.62 & 8.22 & 11.79 & 18.58 & 14.54 \\
\hline
M$_G$/M$_\odot$ & 1.4 & 1.4 & 1.4 & 1.4 & 1.4 & 1.4 \\
\hline
(M$_A$ - M$_G$)/M$_\odot$ & 0.166 & 0.166 & 0.166 & 0.163 & 0.169 & 0.173
\\
\hline
R\,(km) & 12.79 & 12.84 & 12.89 & 12.21 & 10.86 & 11.37 \\
\hline
$\Delta_c$\,(km) & 1.88 & 1.89 & 1.92 & 1.67 & 0.88 & 1.08 \\
\hline
I\,(10$^{44}$ g cm$^2$) & 16.14 & 16.28 & 16.46 & 14.58 & 12.30 & 13.34 \\
\hline
z   & 0.215 & 0.214 & 0.213 & 0.229 & 0.271 & 0.254 \\
\hline
\end{tabular}
\end{center}
\end{table}

\begin{table}
\begin{center}
TABLE VI \\[1cm]
\begin{tabular}{|l|c|c|c|c|c|c|}
\hline
Quantity & RHF1 & RHF6 & RHF8 & RHF9 & RHF12 & RHF13 \\
\hline
\hline
$\epsilon_c~(10^{14}$g/cm$^3$) & 16.3 & 16.8 & 9.3 & 10.1 & 21.8 & 17.8 \\
\hline
P$_c~(10^{34}$dyn/cm$^2$) & 29.52 & 33.59 & 10.09 & 11.91 & 45.2 & 35.47 \\
\hline
M$_G$/M$_\odot$ & 1.4 & 1.4 & 1.4 & 1.4 & 1.4 & 1.4 \\
\hline
(M$_A$ - M$_G$)/M$_\odot$ & 0.21 & 0.25 & 0.16 & 0.16 & 0.216 & 0.219 \\
\hline
R\,(km) & 9.86 & 9.41 & 13.15 & 12.8 & 9.31 & 9.59 \\
\hline
$\Delta_c$\,(km) & 0.8 & 0.81 & 2.4 & 2.25 & 0.88 & 0.98 \\
\hline
I\,(10$^{44}$ g cm$^2$) & 10.43 & 9.86 & 16.11 & 15.16 & 9.25 & 9.80 \\
\hline
z   & 0.312 & 0.335 & 0.208 & 0.215 & 0.341 & 0.325 \\
\hline
\end{tabular}
\end{center}
\end{table}

\begin{table}
\begin{center}
TABLE VII \\[1cm]
\begin{tabular}{|l|c|c|c|c|}
\hline
Quantity & RH1 & RH5 & RHF1 & RHF8 \\
\hline
\hline
$\epsilon_c$\,(MeV/fm$^3$)) & 403.12 & 479.40 & 869.83 & 438.85 \\
\hline
$R_{\rm eq}$ (km)           & 13.72  & 12.95  & 10.14  & 14.49  \\
\hline
$R_{\rm p}$ (km)            & 12.11  & 11.68  &  9.65  & 12.6   \\
\hline
$\log I/({\rm g\,cm}^2)$    & 45.17  & 45.13  & 45.00  & 45.19  \\
\hline
$t$                         & 0.036  & 0.03   & 0.016  & 0.038  \\
\hline
$\beta$                     & 0.658  & 0.646  & 0.572  & 0.672  \\
\hline
$z_{\rm p}$                 & 0.232  & 0.244  & 0.323  & 0.22   \\
\hline
$e$                         & 0.47   & 0.43   & 0.31   & 0.49   \\
\hline
$\Pi$ (km$^3$)              & 7.27   & 5.14   & 1.51   & 8.32   \\
\hline
\end{tabular}
\end{center}
\end{table}

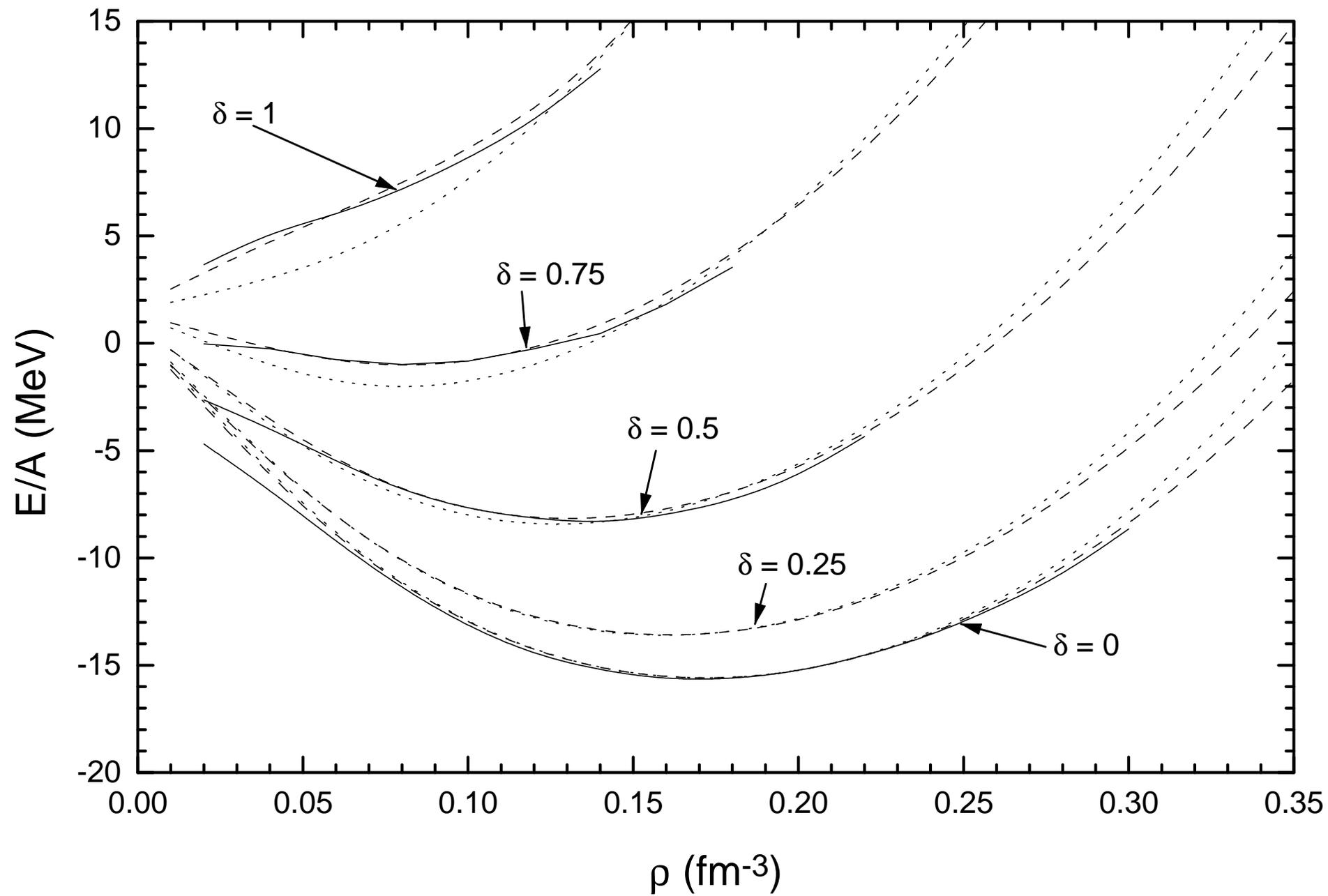
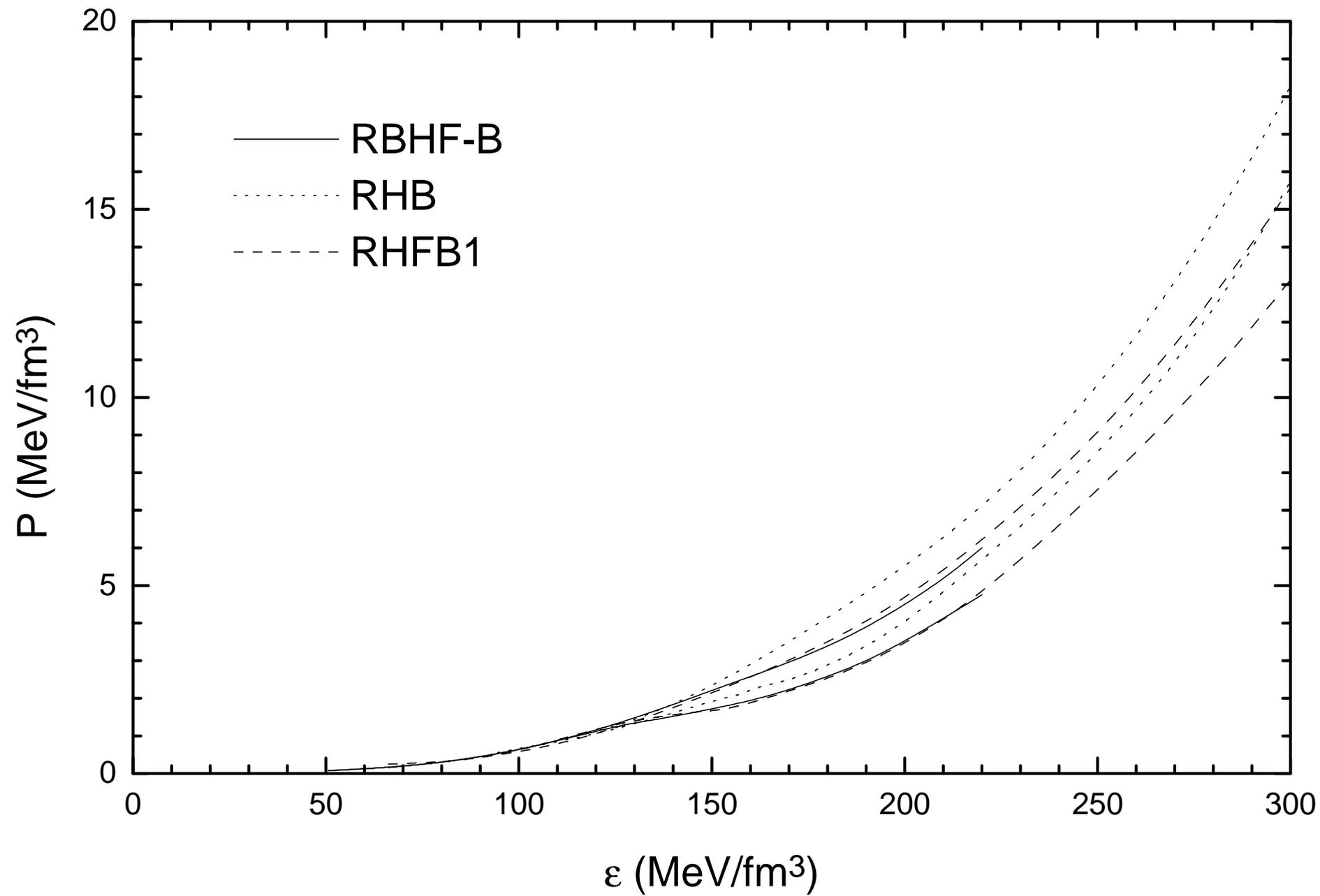
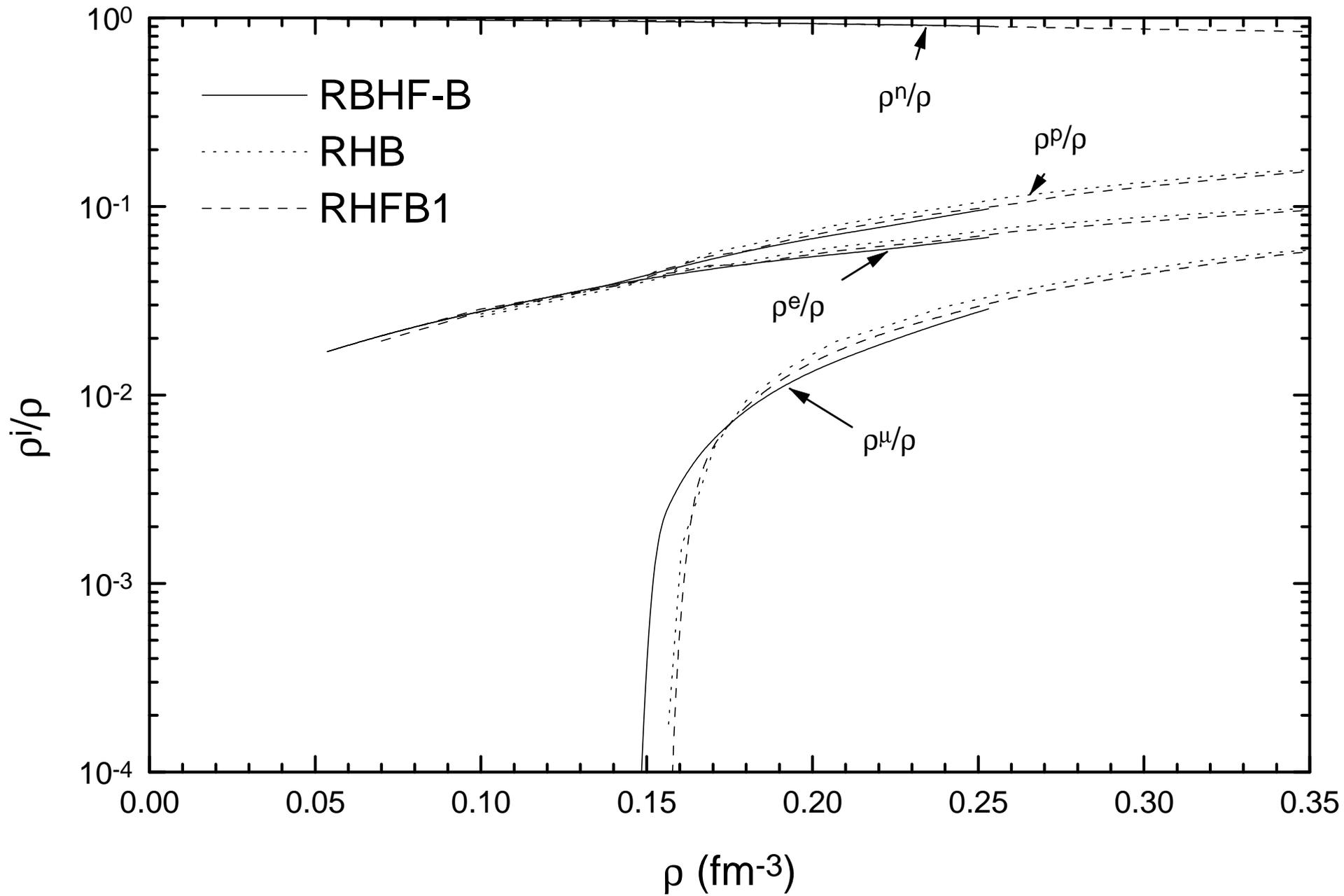
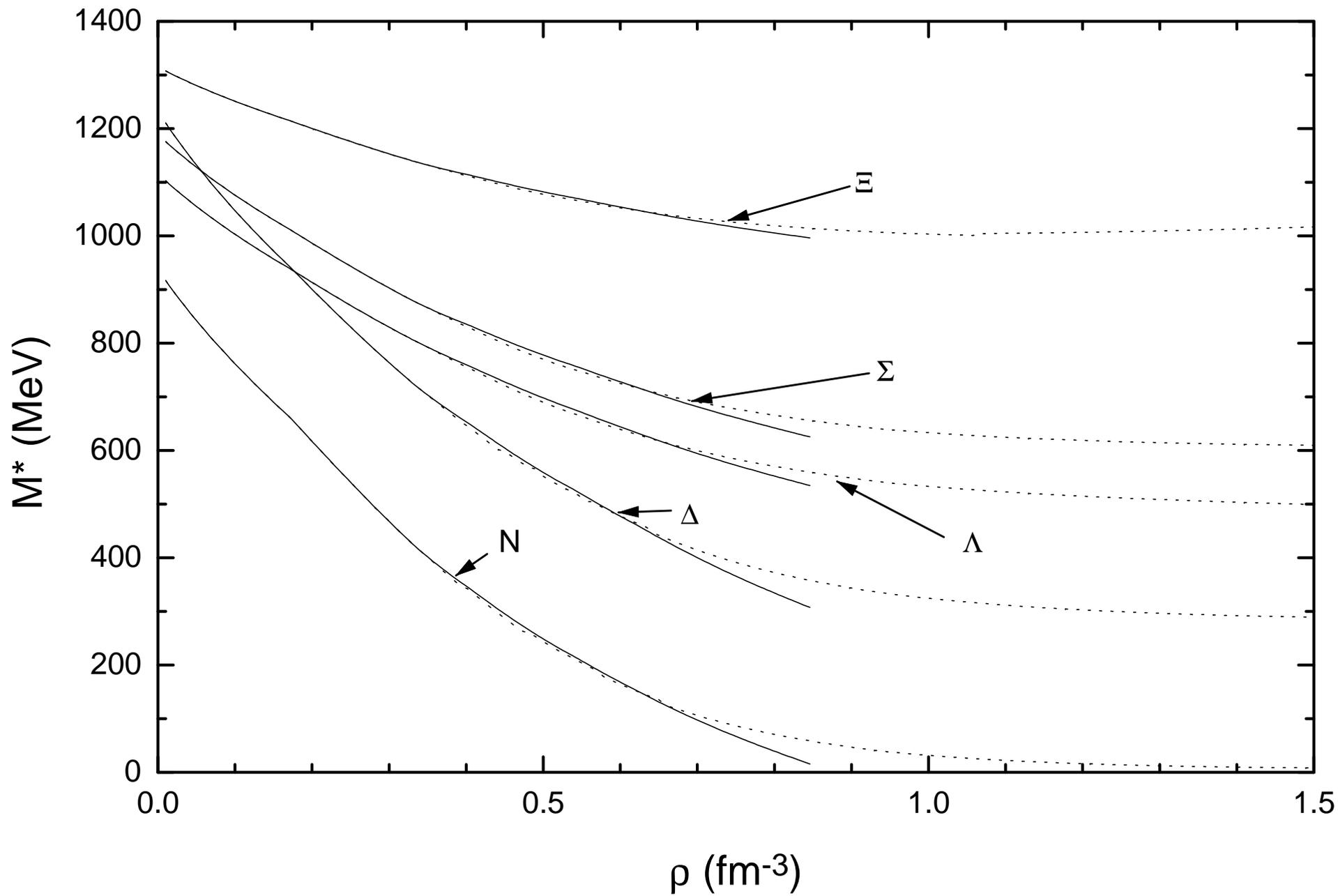
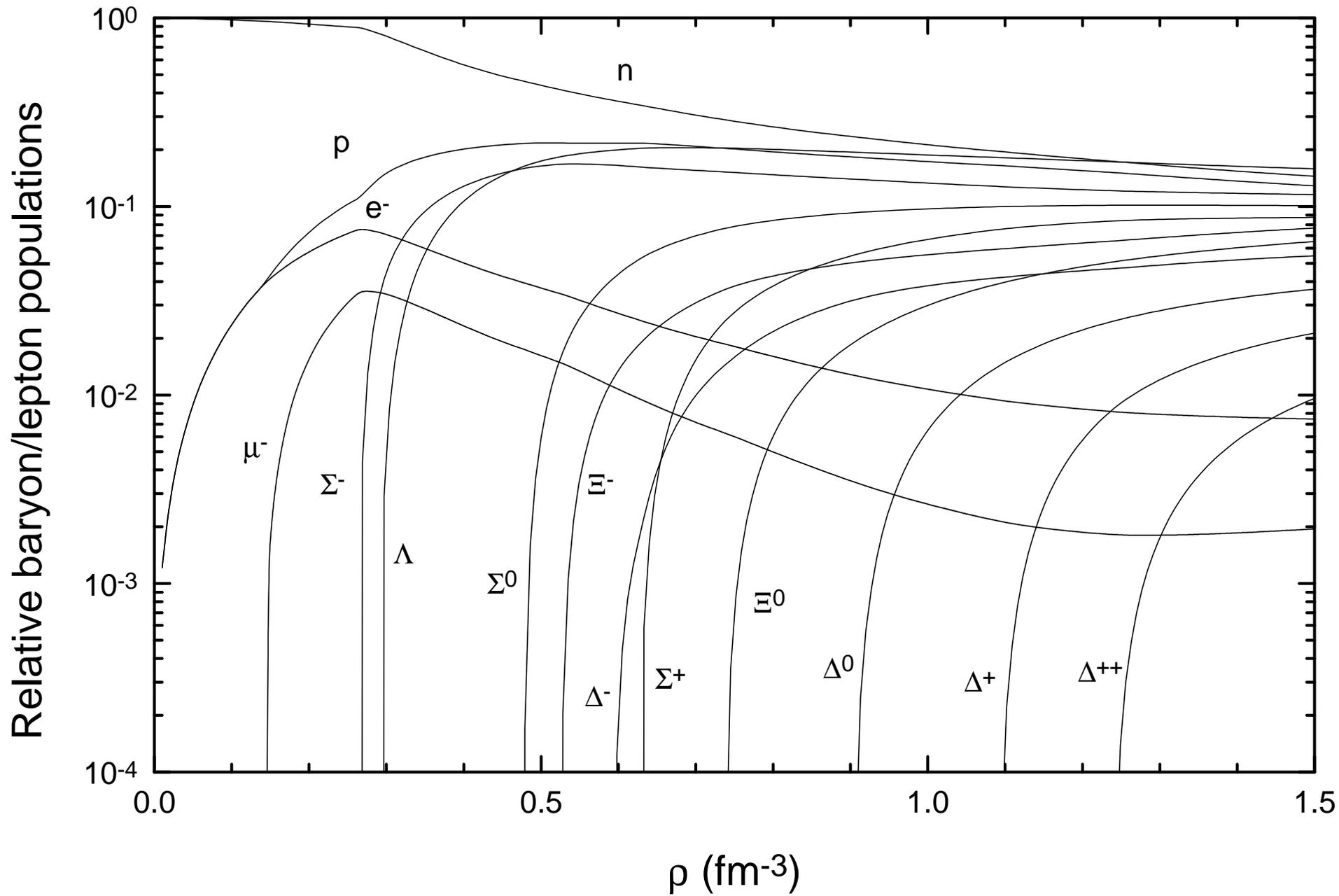
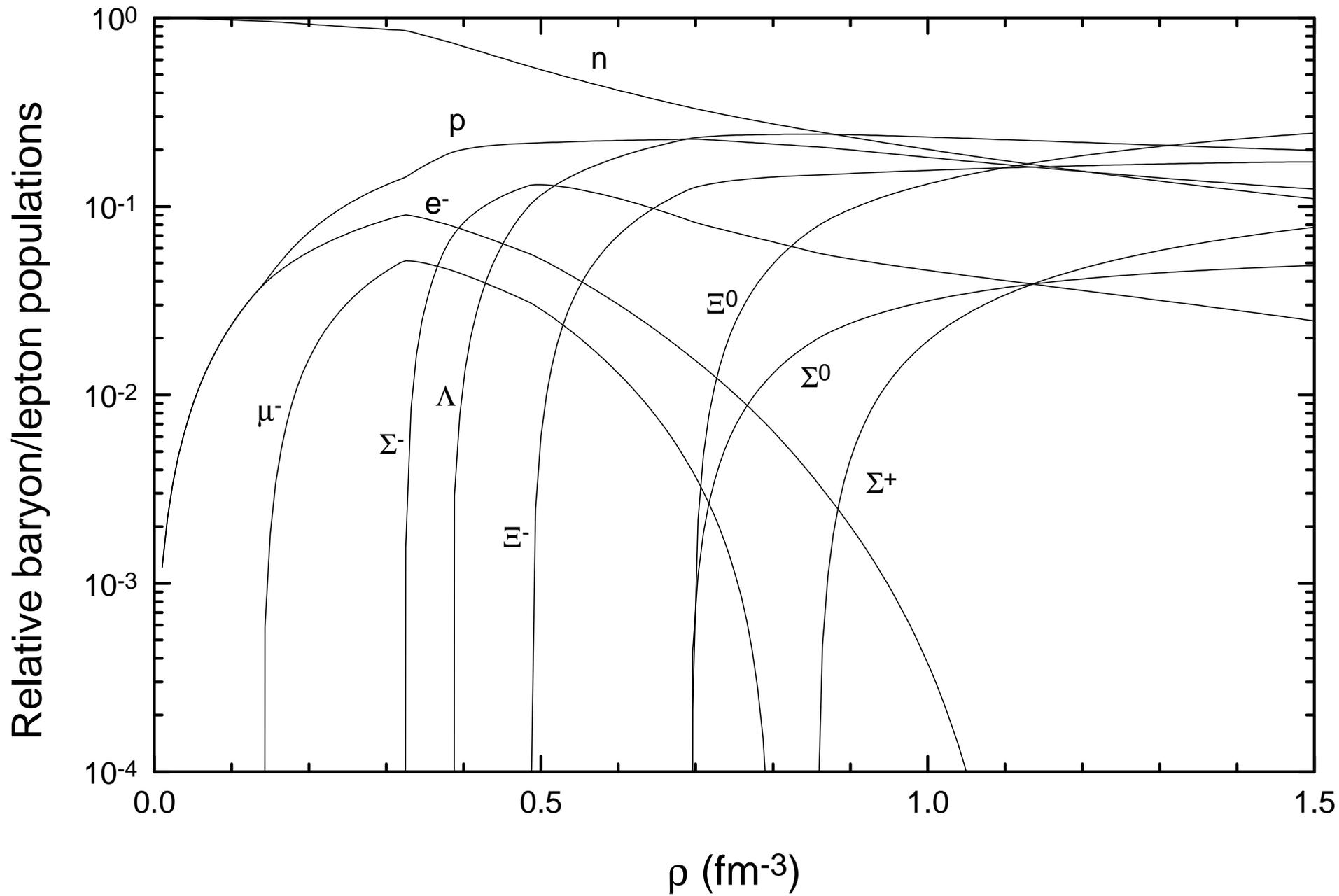
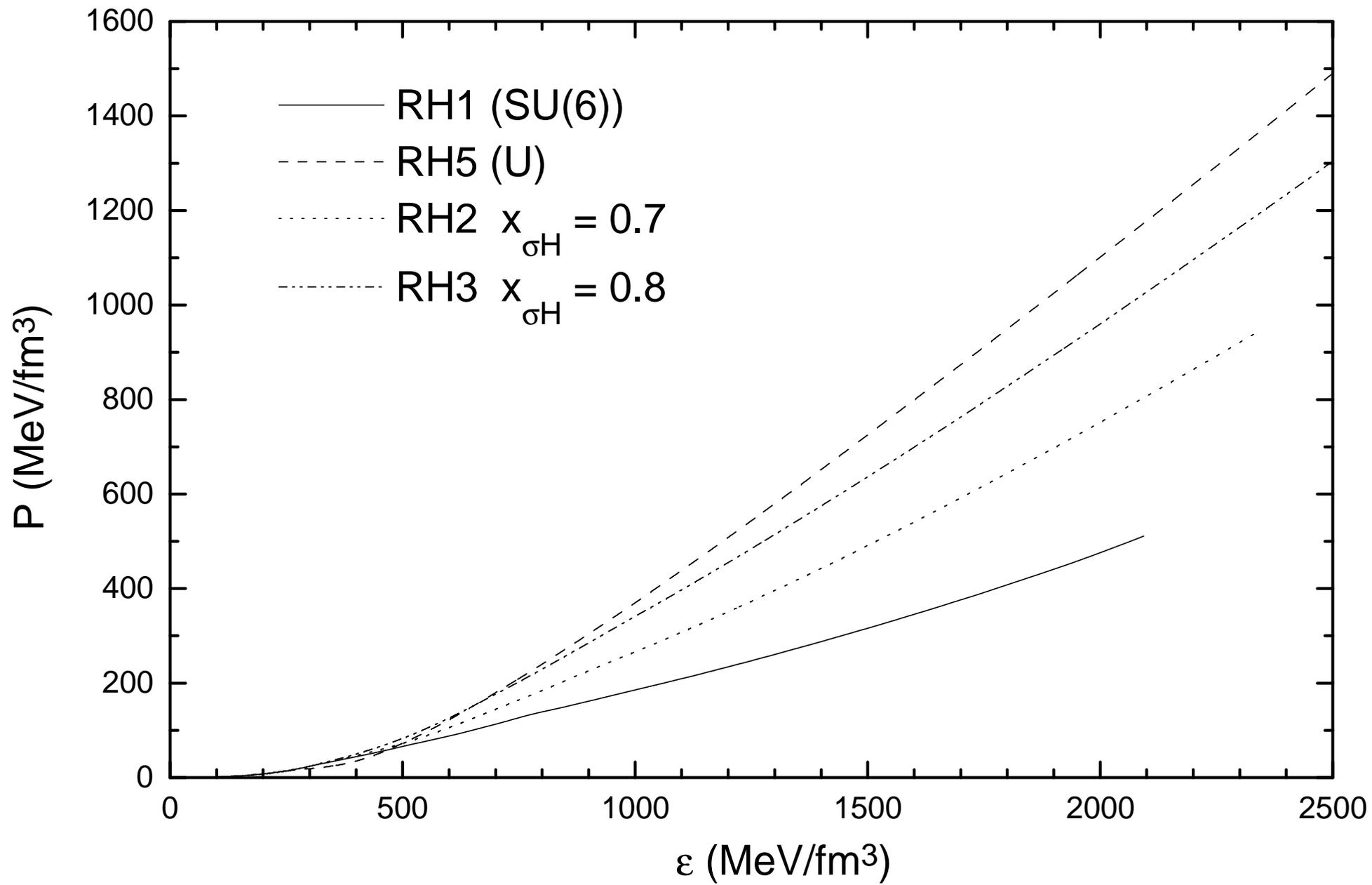
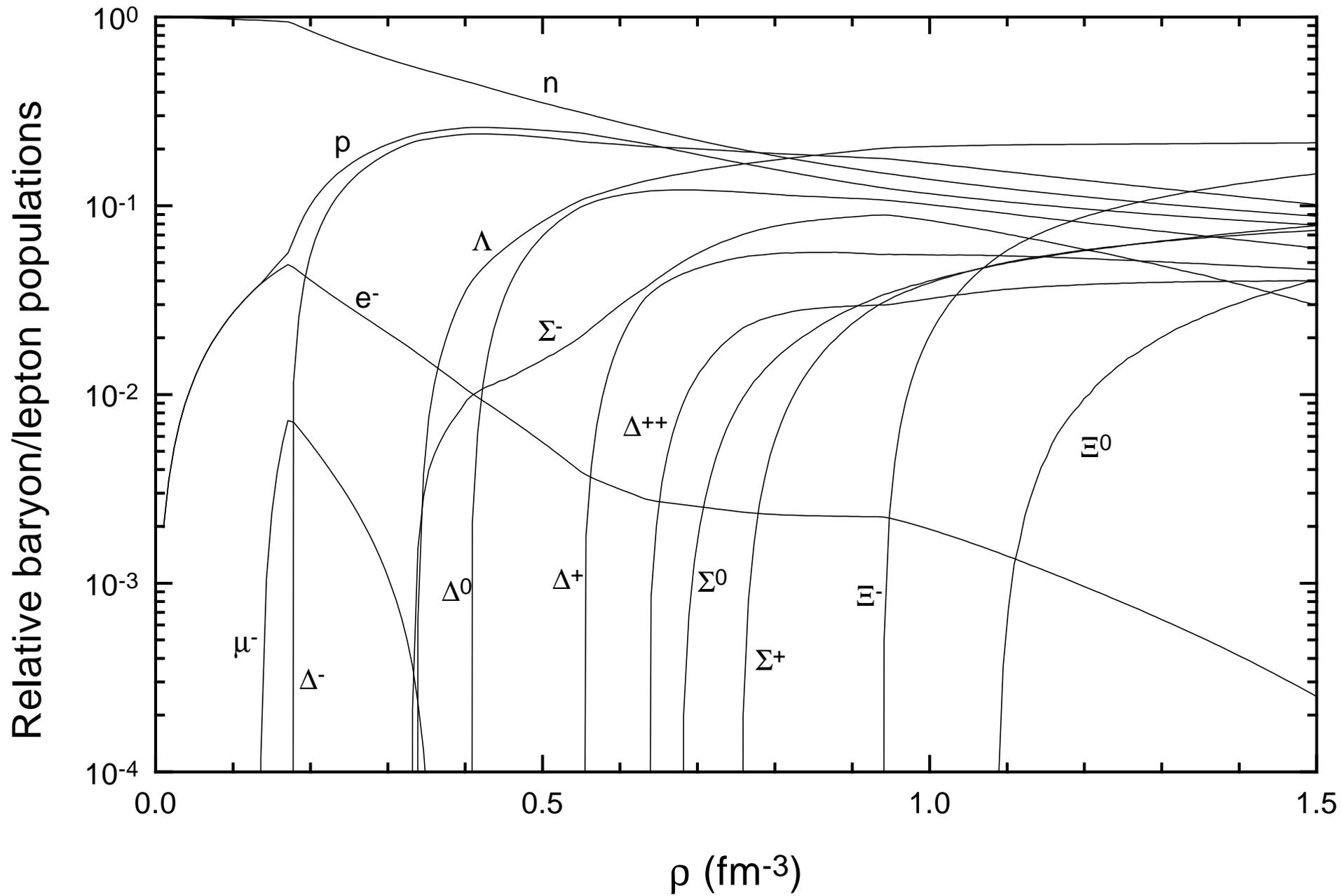
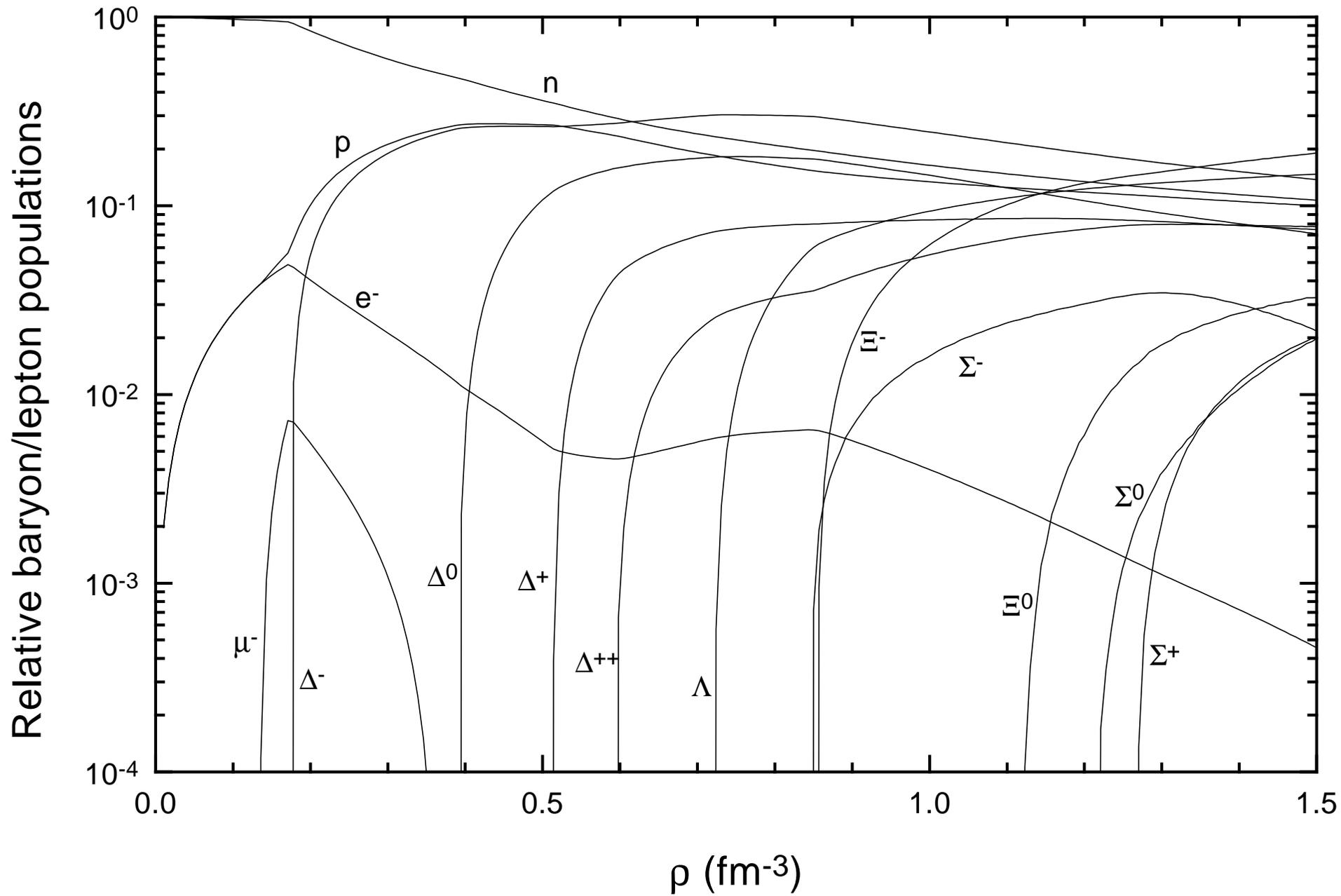
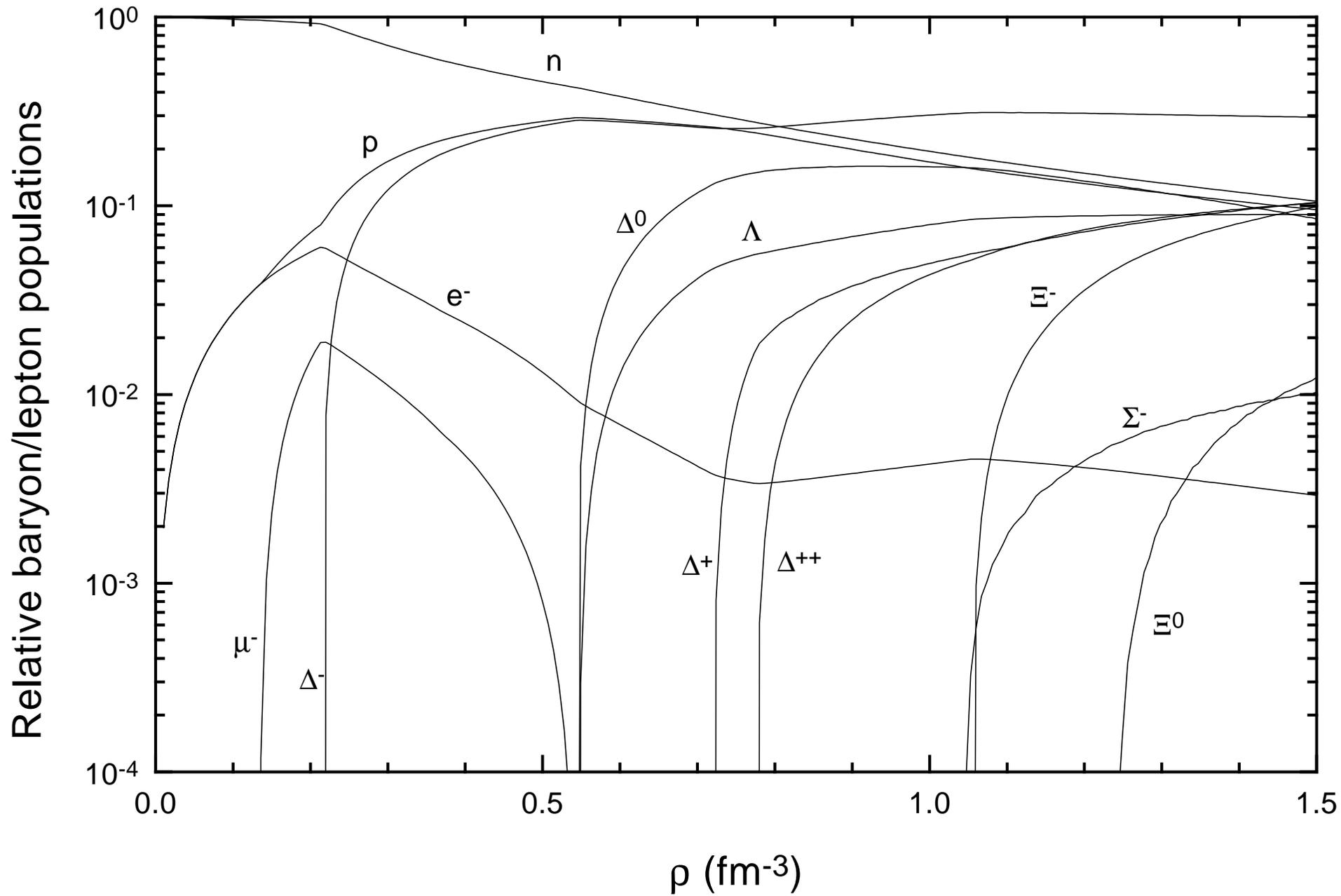
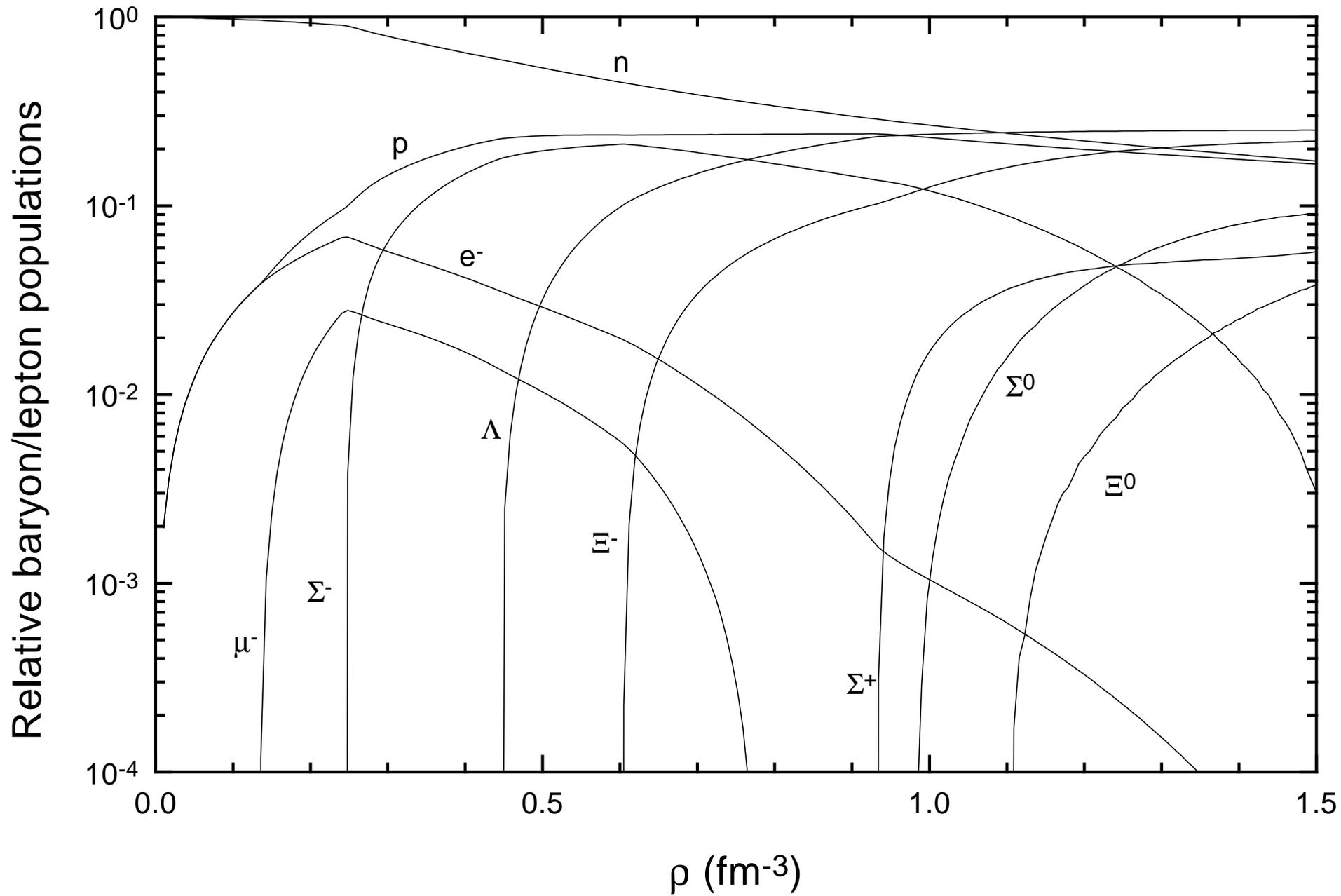
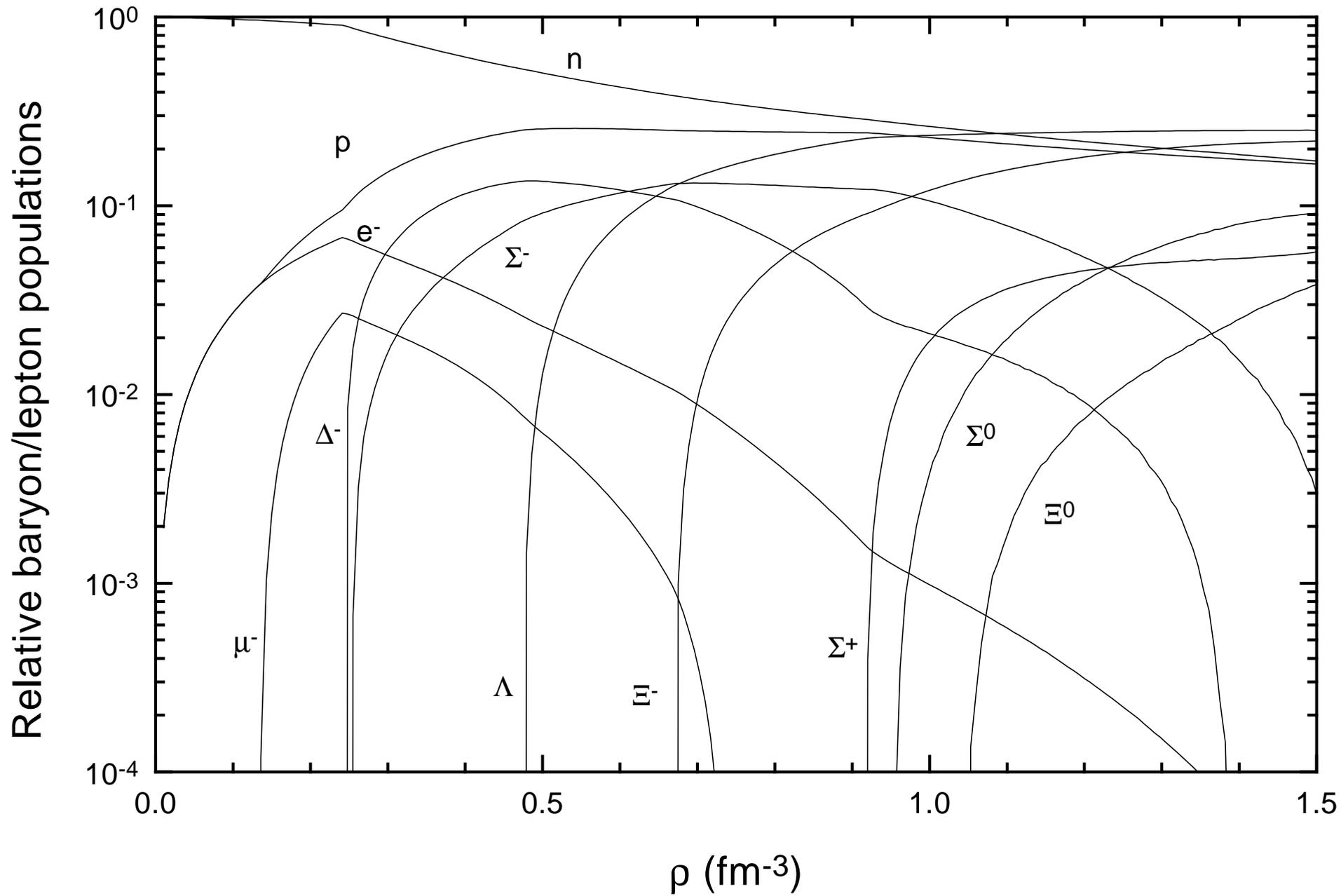
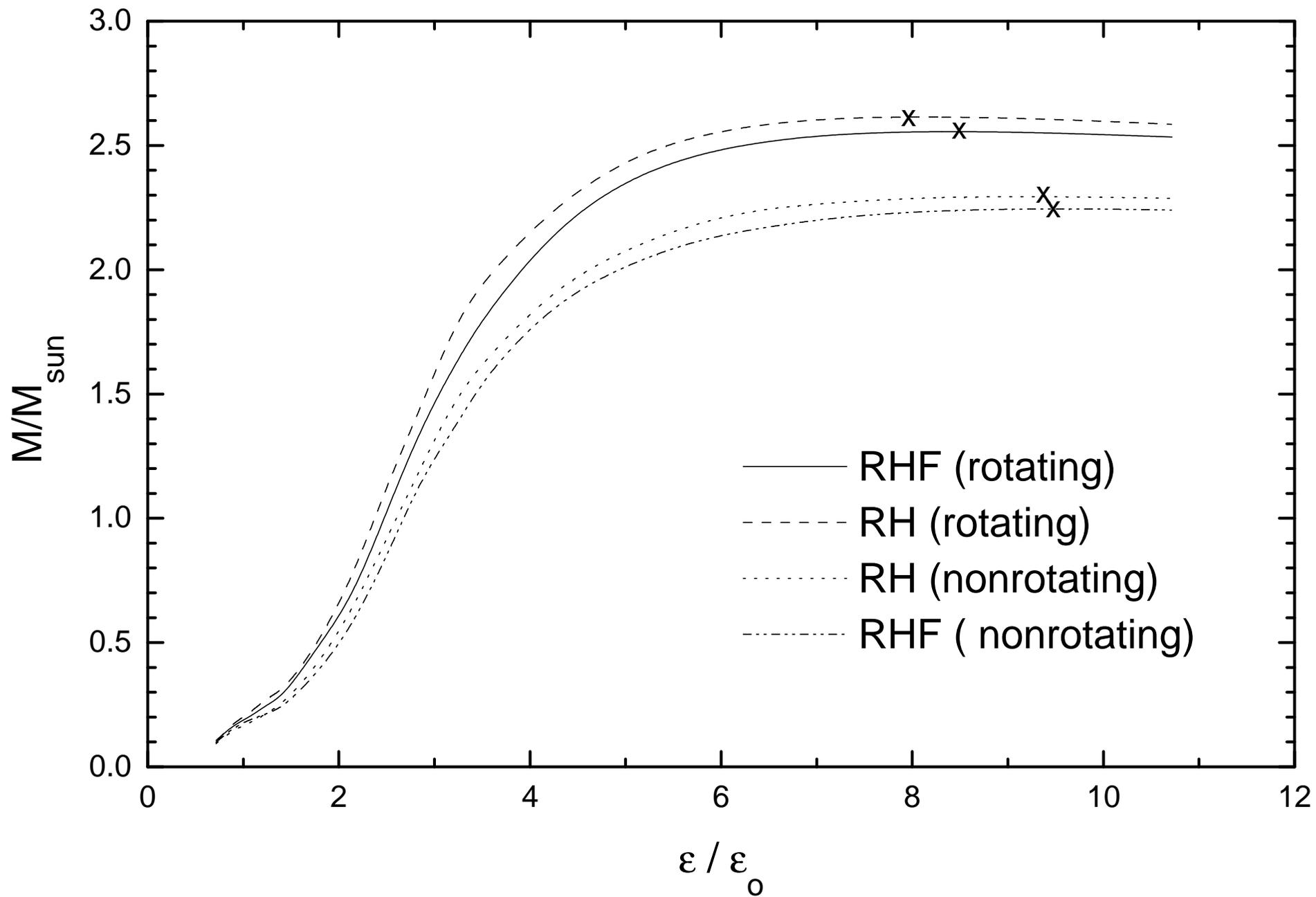
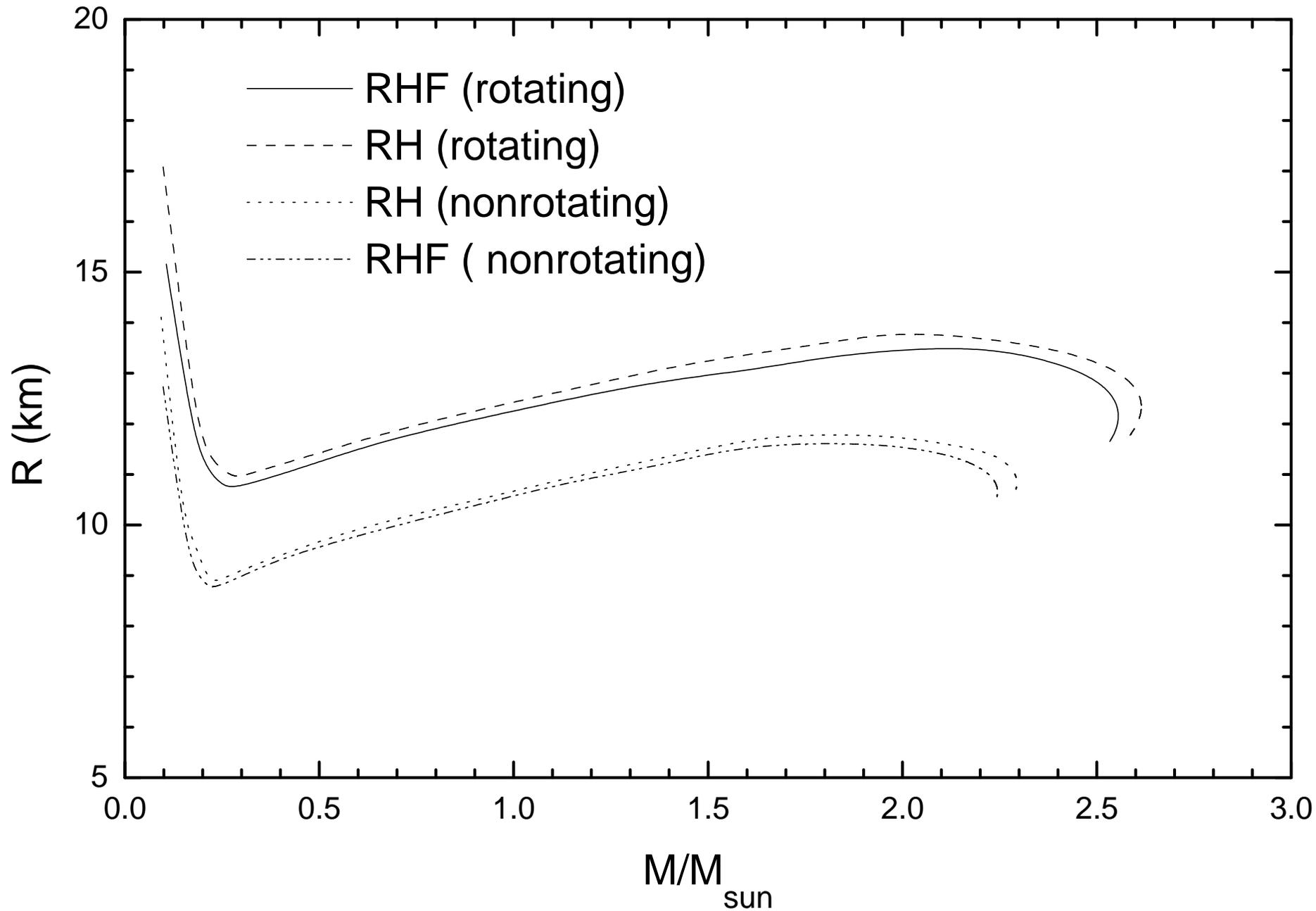
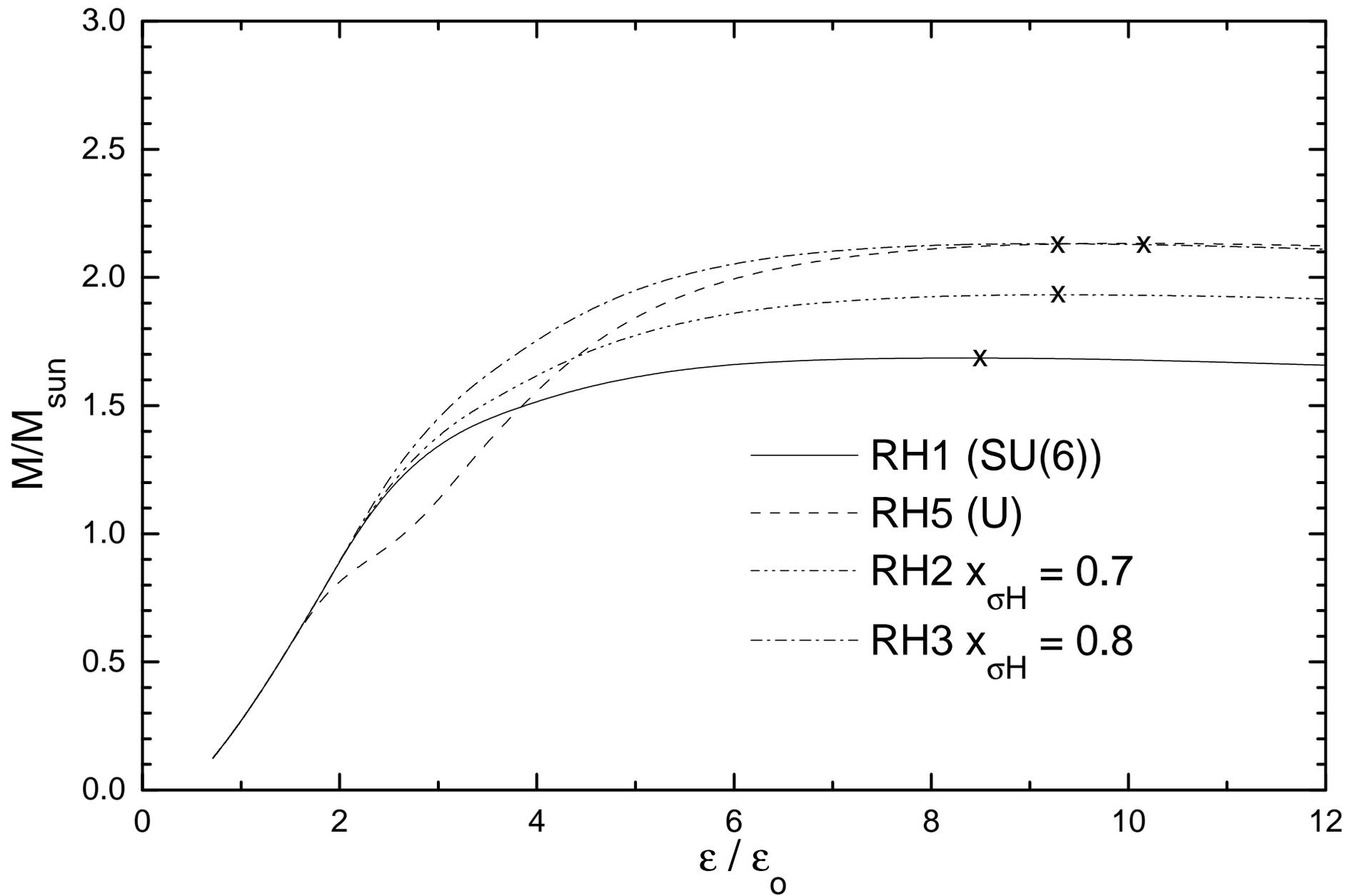
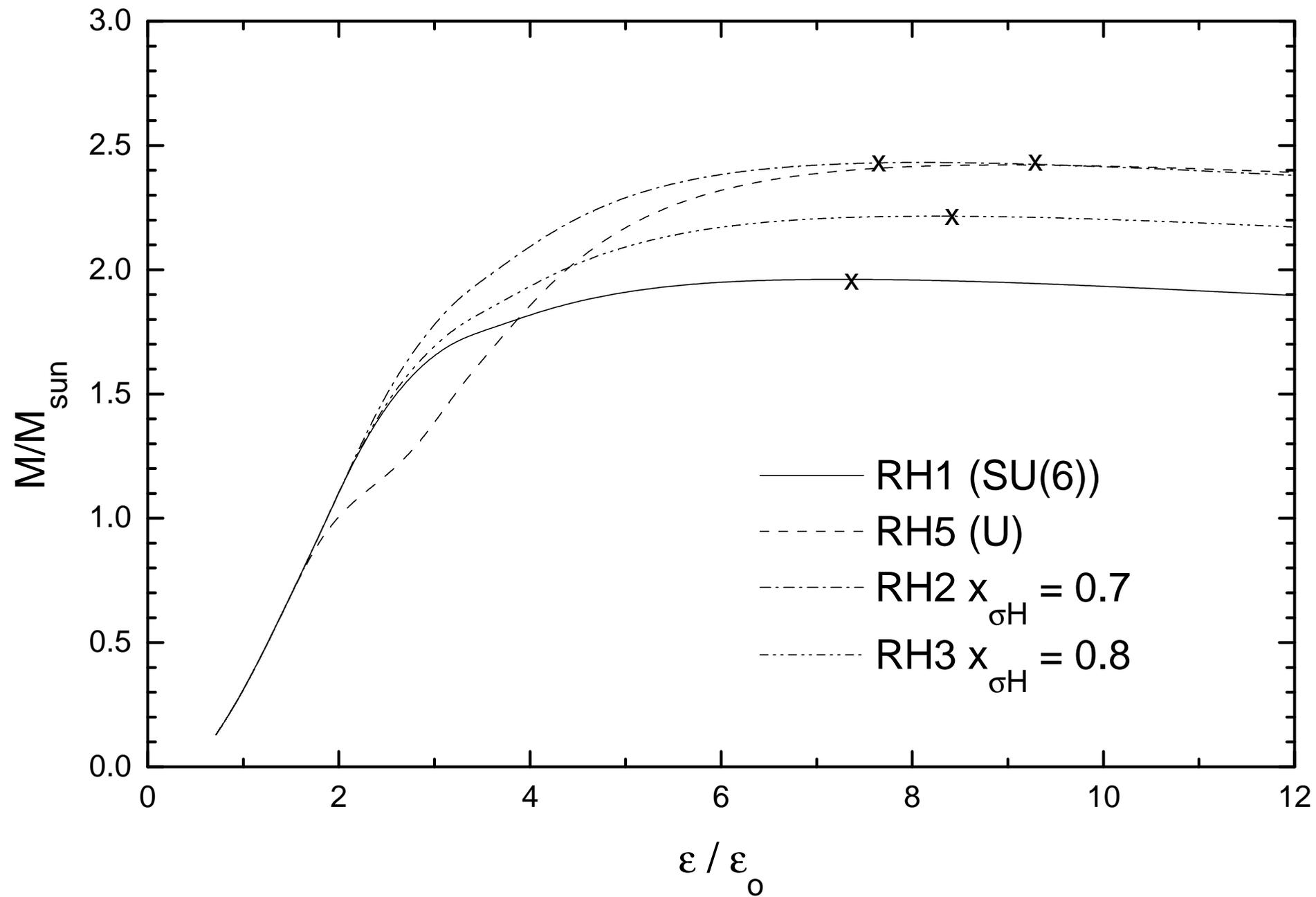
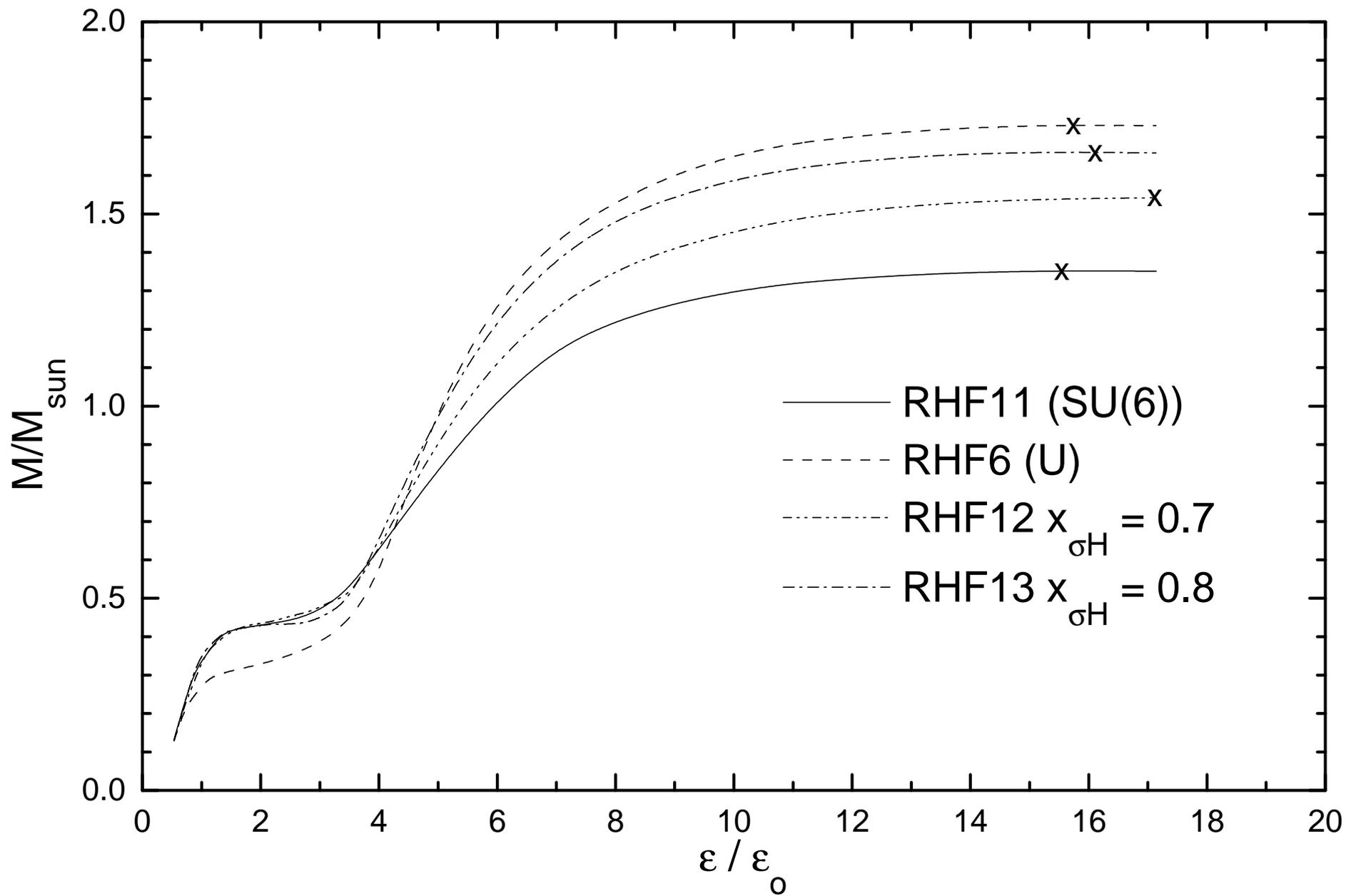
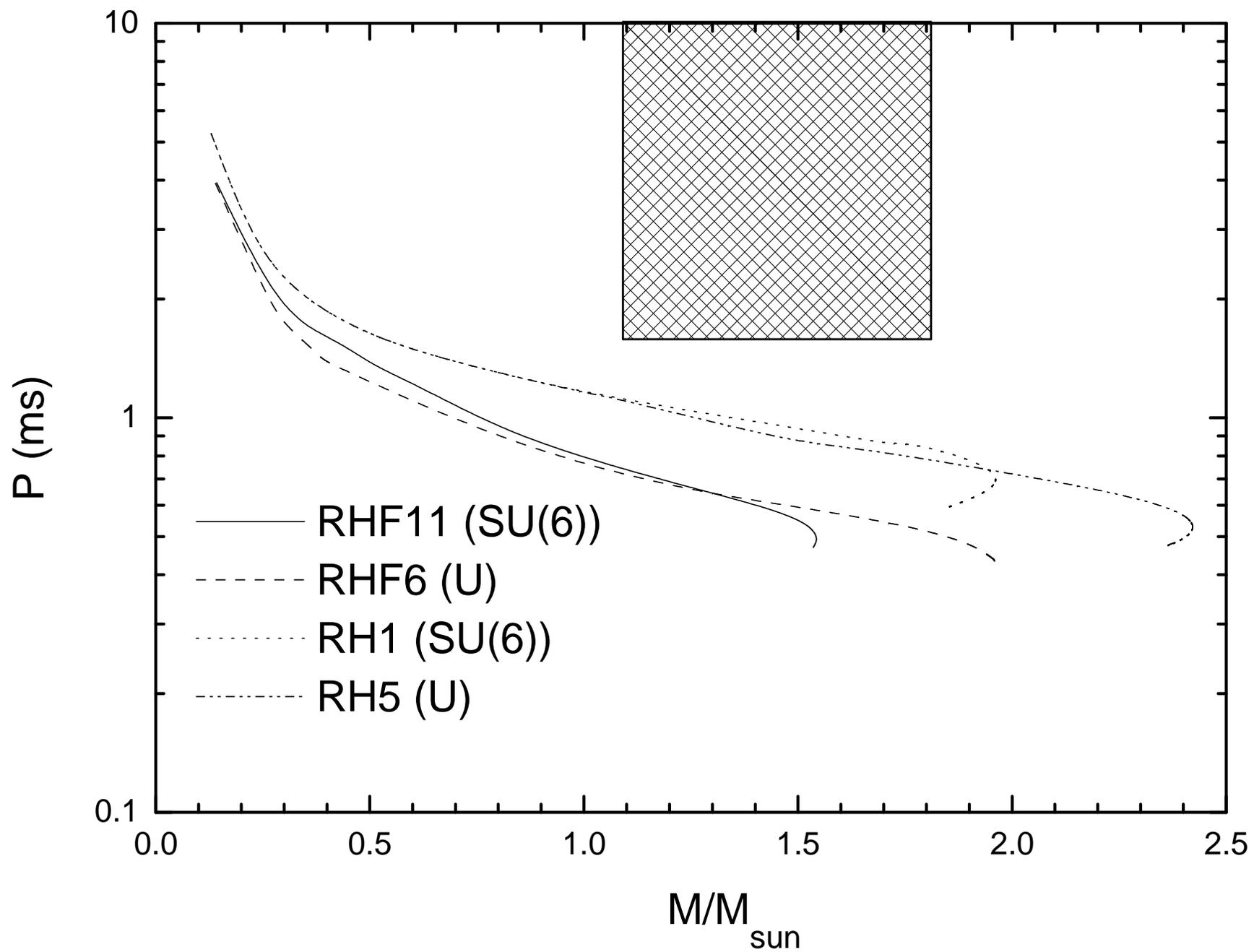
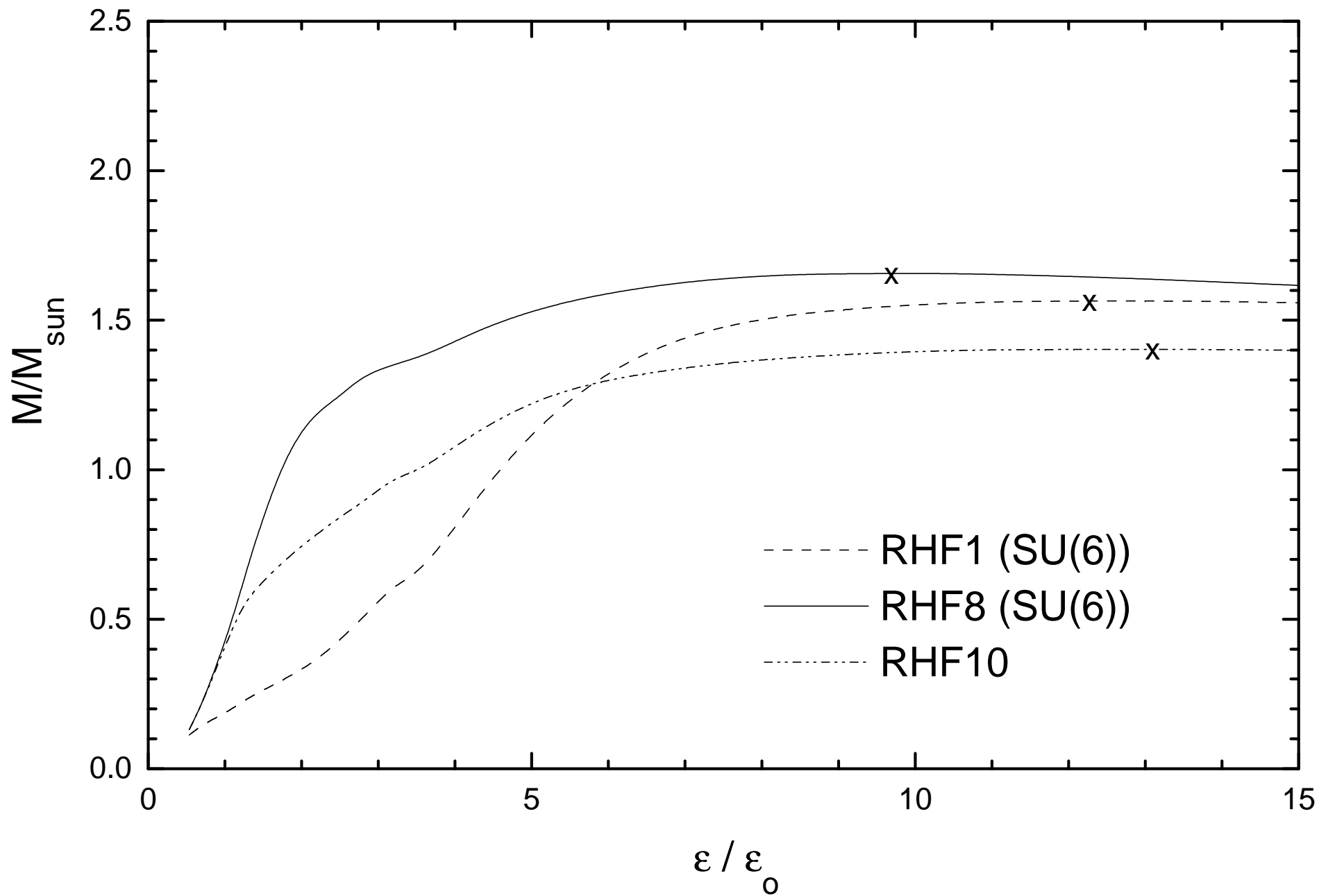
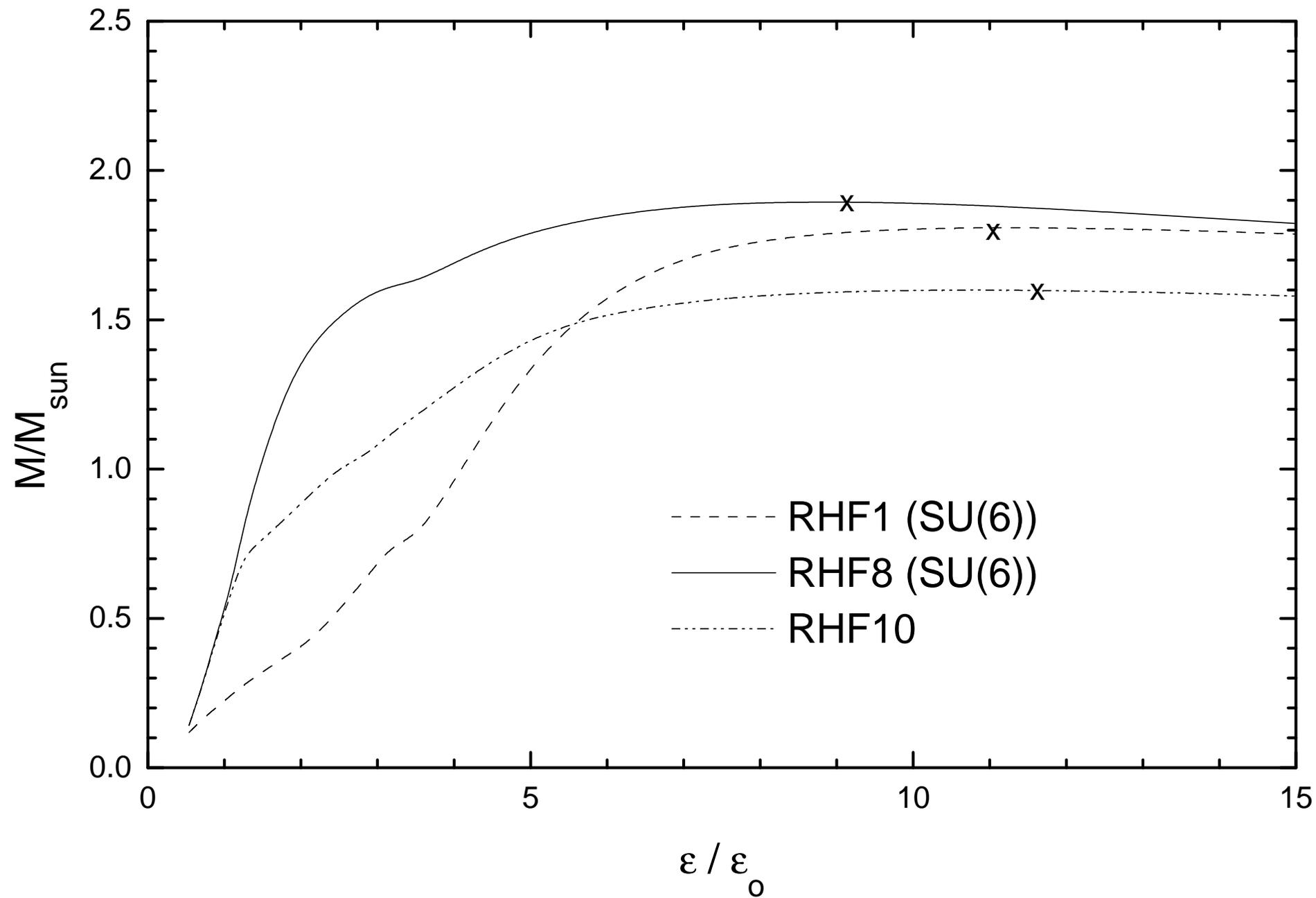
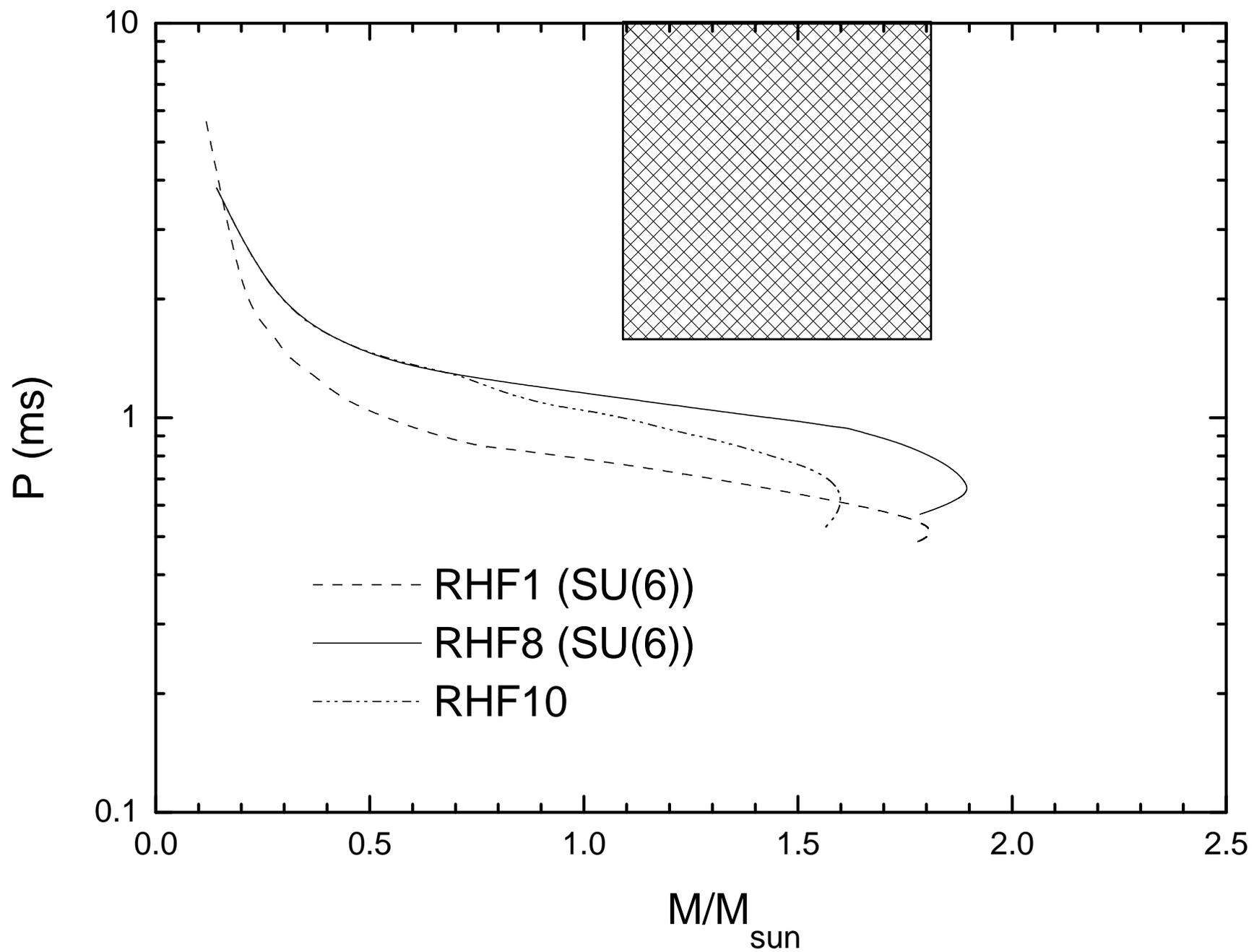
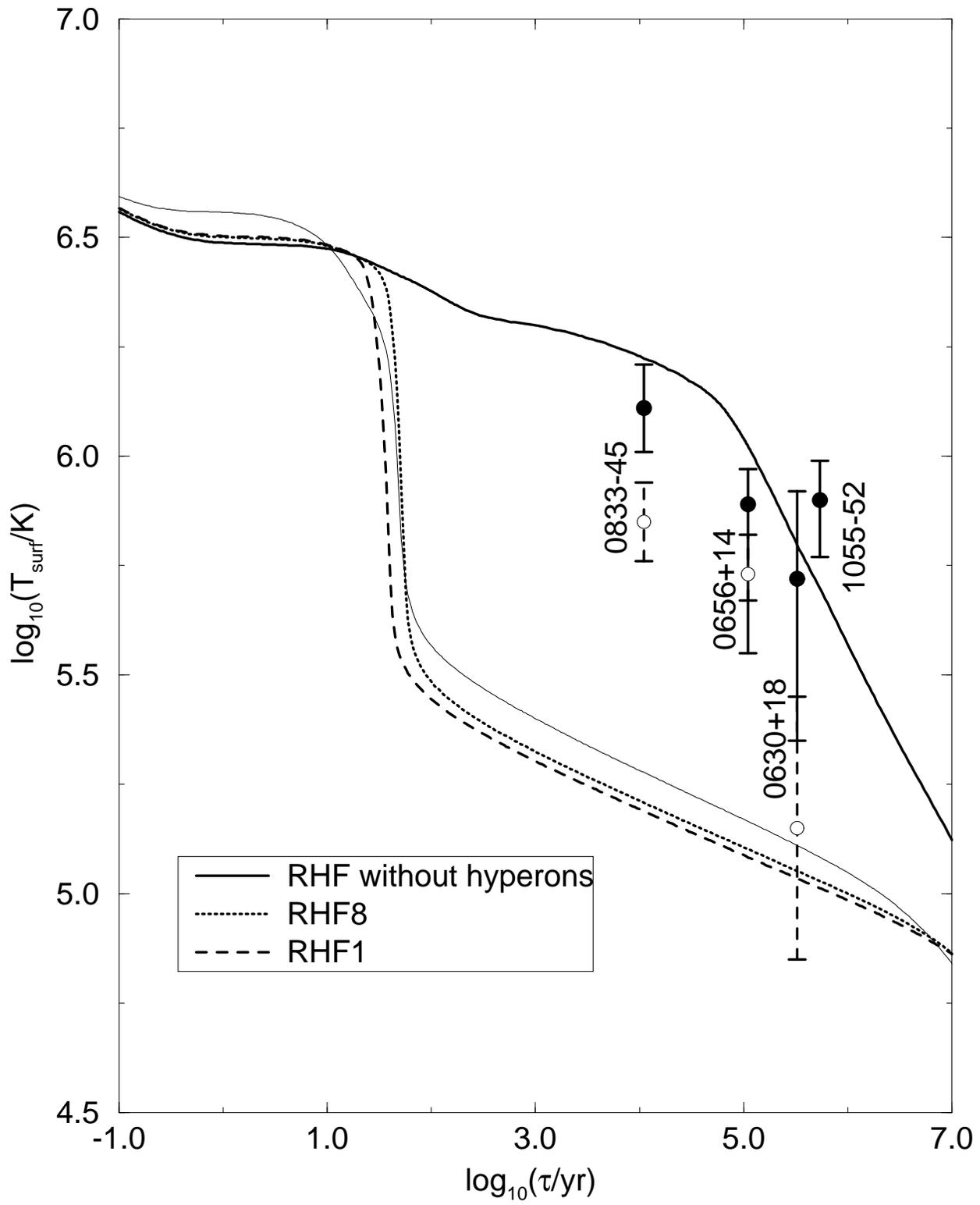
\end{document}